\pdfoutput=1
\documentclass[a4paper,fleqn]{cas-sc}

\usepackage[numbers]{natbib}

\usepackage{mathtools}
\usepackage{tabularx}
\usepackage{subcaption}
\usepackage{algorithm}
\usepackage{algpseudocode}
\usepackage{lineno}

\renewcommand{\ttdefault}{lmtt}

\newcommand{\circled}[1]{\textcircled{\scriptsize #1}}

\ExplSyntaxOn
\cs_set:Npn \__cas_head: { }
\ExplSyntaxOff
\begin{document}

\let\WriteBookmarks\relax
\def\floatpagepagefraction{1}
\def\textpagefraction{.001}

\shorttitle{Elastoplastic inherent strain-based topology optimization}
\shortauthors{T. Miki et~al.}

\title[mode=title]{Elastoplastic inherent strain-based topology optimization for residual stress reduction in metal additive manufacturing}

\author[1]{Takao Miki}
\cormark[1]
\ead{mikit@orist.jp}
\credit{Conceptualization, Methodology, Software, Validation, Investigation, Visualization, Writing -- original draft, Funding acquisition}

\affiliation[1]{organization={Osaka Research Institute of Industrial Science and Technology},
  addressline={7-1, Ayumino-2},
  city={Izumi},
  postcode={594-1157},
  state={Osaka},
  country={Japan}}

\author[2]{Jike Han}
\credit{Methodology, Writing -- review \& editing}

\affiliation[2]{organization={The Hakubi Center for Advanced Research, Kyoto University},
  addressline={Yoshida-honmachi, Sakyo-ku},
  city={Kyoto},
  postcode={606-8501},
  country={Japan}}

\author[3]{Kazuhiro Izui}
\credit{Writing -- review \& editing}

\affiliation[3]{organization={Department of Micro Engineering, Kyoto University},
  addressline={Kyoto-daigaku Katsura C3, Nishikyo-ku},
  city={Kyoto},
  postcode={615-8540},
  country={Japan}}

\author[4]{Shinji Nishiwaki}
\credit{Writing -- review \& editing}

\affiliation[4]{organization={Department of Mechanical Engineering and Science, Kyoto University},
  addressline={Kyoto-daigaku Katsura C3, Nishikyo-ku},
  city={Kyoto},
  postcode={615-8540},
  country={Japan}}

\cortext[1]{Corresponding author.}

\begin{abstract}
This paper proposes a topology optimization method for reducing the residual stress arising in the building process of metal additive manufacturing. First, a layer-by-layer process analysis model based on an elastoplastic inherent strain method is introduced. In this model, the incremental displacement is solved anew at each layer step, and the stress history is explicitly incorporated into the constitutive equation as the stress accumulated up to the previous step, which guarantees the stress continuity across layer interfaces without introducing activation strains. Next, the design sensitivity of this analysis model is derived based on the adjoint method. Taking the pair of the stress and the equivalent plastic strain as the state variables reduces the dependency between layer steps to a one-step recurrence, and the adjoint fields are constructed as a layer-by-layer reverse sweep that reuses the coefficient tensors obtained in the forward analysis. Consequently, the cost of the sensitivity analysis scales linearly with the number of layers and remains of the same order as that of the forward analysis. An optimization problem is then formulated based on the density method to minimize the P-norm of the residual stress at the completion of the building process under the volume and final-use compliance constraints, and the derived sensitivities are verified by comparison with central finite differences. Finally, the proposed method is demonstrated through two- and three-dimensional examples of residual stress minimization under a compliance constraint. The results clarify that, under the elastoplastic analysis, the maximum residual stress is bounded by the yield surface, and the optimization therefore reduces the extent of the yielded and plastic strain accumulating regions rather than the peak stress value.
\end{abstract}


\maketitle

\section{Introduction}\label{sec:1}

Structural optimization is a design methodology that derives optimal structures from a mathematical and mechanical point of view, and it has been widely used in engineering design for purposes such as performance improvement and weight reduction. Among structural optimization methods, topology optimization \cite{BendsoeKikuchi1988,BendsoeSigmund2003} offers a high degree of design freedom, allowing changes not only in the outer shape but also in the number, position, and size of holes, and can create structures that maximize or minimize a prescribed performance measure. Since it can produce innovative high-performance structures that are unattainable by conventional design practice, topology optimization has attracted considerable attention in recent years. On the other hand, the obtained structures often exhibit complex and organic geometries, which can be difficult to fabricate with conventional manufacturing techniques such as machining.

Metal additive manufacturing (AM) has emerged as a promising solution to this issue. In particular, laser powder bed fusion (LPBF) is the most widely used process for metal AM and is expected to enable the fabrication of the complex structures obtained by topology optimization. The integration of topology optimization and AM enables a consistent optimization from design to manufacturing, and the creation of lightweight high-performance components, improvements in manufacturing efficiency, and reductions in cost and energy consumption have been reported in industrial fields such as aerospace, automotive, energy, and heat exchangers \cite{Zhu2016,Bayat2023}.

Metal AM, and LPBF in particular, nevertheless involves several manufacturing issues that restrict the freedom of design and fabrication. A representative issue is the residual stress accumulated during the building process and the resulting process-induced distortion. In LPBF, local melting by a laser and rapid solidification of metal powder are repeated layer by layer. In this layer-wise building process, nonuniform strains caused by phase transformation and by the thermal expansion and contraction that accompany local temperature changes accumulate under the constraint of the underlying layers and surrounding regions, and a residual stress field is formed accompanied by plastic deformation. As a result, warpage, distortion, and even cracking may occur, which can severely degrade the dimensional accuracy and reliability of the fabricated parts.

Another important restriction is the limitation on the build angle, that is, the overhang. In LPBF, since the powder beneath a layer remains unmelted, downward-facing surfaces close to horizontal are not self-supporting, and support structures become inevitable. This not only restricts the design freedom but also increases the burden and cost of post-processing for support removal. In addition, for structures with complex internal cavities and channels, unmelted powder remains after fabrication, and powder removal itself becomes a design constraint. Metal AM thus involves both physical issues inherent in the thermo-mechanical process and geometric or post-processing constraints, and design and optimization methods that account for them are required.

Against this background, topology optimization considering the manufacturability of AM has been actively studied. In particular, geometric constraint approaches incorporating overhang constraints have been widely investigated \cite{GaynorGuest2016,Langelaar2017,Qian2017,Allaire2017}. These methods evaluate the overhang angle, that is, the angle between the surface normal and the building direction, and ensure that the structure is self-supporting through smooth boundary control in level set-based methods or through gradient-based constraints in density-based methods, thereby contributing to the reduction of support material and post-processing cost. However, geometric constraint approaches alone cannot address the physical manufacturing issues such as residual stress and distortion.

In parallel, studies that integrate process analysis models of AM with optimization methods have been conducted to suppress residual stress and distortion at the design stage. Representative approaches include thermo-mechanical coupled analysis \cite{AllaireJakabcin2018} and the inherent strain method \cite{Ueda1975,KellerPloshikhin2014,Liang2018,Bellet2023}, which have been applied to topology optimization problems considering residual stress distributions and distortion. The former requires coupled nonlinear analyses and is computationally expensive, whereas the latter is an elastic or elastoplastic analysis that employs strain fields identified by experiments or high-fidelity analyses and can predict part-scale residual stress and distortion at low cost. For example, Xu et al.~\cite{Xu2022} evaluated the layer-wise stress field by a simplified inherent strain model and proposed a density-based optimization method that simultaneously considers residual stress and overhang constraints. Miki and Yamada \cite{MikiYamada2021} and Miki \cite{Miki2023} constructed level set-based topology optimization methods that simultaneously consider overhang constraints and AM-induced distortion, and verified their effectiveness with two- and three-dimensional numerical examples.

Although these previous studies have enabled structural designs that account for the physical issues specific to AM to some extent, the analysis models employed are simplified elastic analyses, and the influence of plastic deformation, which is dominant in the actual LPBF process, has not been fully evaluated.

Outside the context of AM, topology optimization with elastoplastic constitutive laws has been studied since the late 1990s \cite{Maute1998,Schwarz2001}. It has since been applied to energy absorption and dissipation \cite{Kato2015,NakshatralaTortorelli2015,Wallin2016}, to stress constraints \cite{Amir2017}, and to large-scale problems \cite{Granlund2024}. Because the response of such structures is path dependent, the design sensitivity cannot be recovered from the terminal state alone. It is instead evaluated by an adjoint scheme that runs backward through the load steps, and general formulations of this path-dependent sensitivity analysis are now well established \cite{Michaleris1994,Alberdi2018,Han2025}. Approximations that avoid the path dependence altogether have also been proposed \cite{LiWallin2024}. The same history dependence arises in the layer-by-layer building process of AM, where the layer index plays the role of the load step.

Dugast and To~\cite{DugastTo2023} introduced an analysis model based on an elastoplastic constitutive law capable of representing plastic deformation and proposed a method for optimizing support structures. However, since their analysis model is based on activation strains, the displacement field at the moment of element activation is carried over into the constitutive equation as the activation strain, which introduces additional dependencies between layer steps; the sensitivity analysis therefore becomes complicated and the computational cost increases. This limits its applicability to the optimization of high-resolution three-dimensional models with product performance constraints. Consequently, an optimization method that introduces an elastoplastic constitutive law to represent plastic deformation while retaining the overall computational efficiency including the sensitivity analysis is strongly demanded.

To resolve the above issues, in this study we construct an elastoplastic topology optimization method aimed at reducing the residual stress arising in metal AM. First, to accurately evaluate the plastic deformation accumulated in a part during the building process, we introduce a layer-by-layer process analysis model that is based on an elastoplastic constitutive law and does not use activation strains. The state fields of each layer are constructed recursively by explicitly incorporating the stress accumulated up to the previous layers into the constitutive equation, which guarantees the stress continuity across layer interfaces. This allows the history-dependent stress distribution and the accumulation of plastic strain, which cannot be captured by conventional elastic models, to be reproduced accurately. Next, as the sensitivity analysis, we propose an adjoint analysis method in which the adjoint variables are carried over layer by layer in the reverse order of the building sequence. Owing to the formulation without activation strains, the reverse update of the adjoint fields can be constructed by reusing the coefficient tensors obtained in the return mapping of the forward analysis. That is, the adjoint fields can be constructed with the same layer-by-layer structure as the physical fields, so that the implementation of the adjoint analysis becomes analogous to that of the forward analysis, and the memory usage and implementation effort are substantially reduced. The derived sensitivities are verified by comparison with finite differences, and the effectiveness of the proposed method is demonstrated by two- and three-dimensional numerical examples that minimize the residual stress under a compliance constraint.

By integrating the elastoplastic layer-by-layer process analysis model with the adjoint sensitivity analysis that exploits the layer-by-layer structure of the building process, the proposed method realizes a high-fidelity structural optimization that faithfully reflects the manufacturing process, thereby aiming to improve product reliability while ensuring product performance.

The remainder of this paper is organized as follows. Section~\ref{sec:2} presents the layer-by-layer AM process analysis model based on the elastoplastic inherent strain method and confirms its behavior with a numerical example. Section~\ref{sec:3} formulates the topology optimization problem that minimizes the residual stress under a compliance constraint based on the density method. Section~\ref{sec:4} derives the design sensitivities using the adjoint method. Section~\ref{sec:5} describes the implementation of the optimization algorithm and the verification of the sensitivities by finite differences. Section~\ref{sec:6} presents two- and three-dimensional numerical examples to validate the effectiveness of the proposed method. Finally, Section~\ref{sec:7} provides the conclusions of this paper.

\section{Layer-by-layer AM process analysis model based on the elastoplastic inherent strain method}\label{sec:2}

In this section, we formulate a layer-by-layer process analysis model based on the elastoplastic inherent strain method for evaluating the residual stress and plastic strain accumulated in a part during the layer-wise building process of metal AM. Section~\ref{sec:2-1} outlines the basic concept of the inherent strain method, its extension to layer-by-layer AM analysis, and the lineage of its elastoplastic variants. Section~\ref{sec:2-2} introduces the quiet element method for modeling layer activation and describes the modeling strategy of this study, which does not use activation strains. Section~\ref{sec:2-3} formulates the elastoplastic constitutive equation that explicitly incorporates the carry-over of the accumulated stress and clarifies the recursive structure of the state fields. Section~\ref{sec:2-4} presents the weak form of the governing equations and the layer-by-layer analysis algorithm, and Section~\ref{sec:2-5} confirms the behavior of the analysis model with a numerical example.

\subsection{Inherent strain method}\label{sec:2-1}

The inherent strain method was proposed by Ueda et al.~\cite{Ueda1975} as an efficient technique for predicting the residual stress and distortion caused by welding processes. Residual stress originates from inelastic strains, such as plastic and thermal strains, generated in the process of local melting and solidification of metallic materials. The inherent strain method identifies the total of these inelastic strains as the inherent strain and applies it as an initial strain in a static mechanical analysis at room temperature, thereby reproducing the residual stress and distortion without performing a transient thermo-mechanical coupled analysis \cite{Bellet2023}. Decomposing the total strain $\boldsymbol{\varepsilon}$ after cooling into elastic, plastic, and thermal components, the inherent strain $\boldsymbol{\varepsilon}^*$ is defined as
\begin{equation}\label{eq:2-1}
\boldsymbol{\varepsilon}^* := \boldsymbol{\varepsilon} - \boldsymbol{\varepsilon}^e = \boldsymbol{\varepsilon}^p + \boldsymbol{\varepsilon}^{th}
\end{equation}
where $\boldsymbol{\varepsilon}^e$, $\boldsymbol{\varepsilon}^p$, and $\boldsymbol{\varepsilon}^{th}$ represent the elastic strain, the plastic strain, and the thermal strain, respectively; $\boldsymbol{\varepsilon}^{th}$ may include volume changes originating from phase transformation.

Since the building process of AM can be regarded as a repetition of welding processes, the inherent strain method has been extended to metal AM by way of its application to multi-pass welding \cite{KellerPloshikhin2014}: the part is divided into multiple layers along the building direction, and the mechanical equilibrium at room temperature is solved sequentially while the inherent strain is applied to each newly added layer, which predicts the part-scale accumulation of residual stress and distortion at low cost. The early inherent strain methods were based on a linear elastic constitutive law, in which the stress grows beyond the yield stress without control and the residual stress is severely overestimated \cite{Bellet2023}; nonlinear variants employing an elastoplastic constitutive law have therefore been proposed \cite{PrabhuneSuresh2020,DugastTo2023}. The analysis model formulated in this section belongs to this elastoplastic lineage.

The inherent strain is identified either by high-fidelity analyses resolving the scanning process, such as the modified inherent strain method \cite{Liang2018,Bellet2023}, or by experiments in which a cantilever-shaped specimen is partially cut and its deflection is measured \cite{MikiYamada2021}. The identification itself is beyond the scope of this study; hereafter, following the common practice in previous studies \cite{Xu2022,DugastTo2023}, an identified uniform constant tensor $\boldsymbol{\varepsilon}^*_0$ is applied at the moment of activation of each macro-layer, an analysis layer thicker than the actual powder layers, into which the part is divided along the building direction.

\subsection{Layer activation and quiet element formulation}\label{sec:2-2}

\begin{figure}[pos=!htbp]
  \centering
  \includegraphics[width=0.60\linewidth]{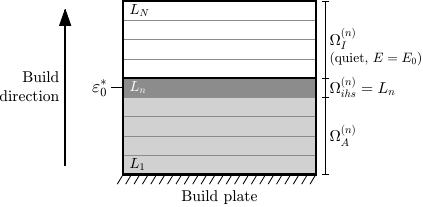}
  \caption{Concept of layer activation and domain decomposition. At step $n$, the fixed domain $\Omega$ is partitioned into the activated domain $\Omega_A^{(n)}$ that has already been built, the layer $\Omega_{ihs}^{(n)} = L_n$ that is activated at this step and subjected to the inherent strain $\boldsymbol{\varepsilon}^*_0$, and the inactive domain $\Omega_I^{(n)}$. The inactive domain is treated as quiet elements with a very small Young's modulus $E_0$.}
  \label{fig:2-1}
\end{figure}

To apply the inherent strain method to the building process of AM, the fixed domain $\Omega \subset \mathbb{R}^d$ ($d = 2, 3$) occupied by the manufactured part is divided into $N$ macro-layers $L_1, L_2, \ldots, L_N$ of constant thickness along the building direction (Fig.~\ref{fig:2-1}):
\begin{equation}\label{eq:2-2}
\Omega = \bigcup_{n=1}^{N} L_n, \qquad L_i \cap L_j = \emptyset \quad (i \neq j)
\end{equation}
The building process is represented by $N$ layer steps, where exactly one macro-layer is activated per step, that is, layer $L_n$ is newly activated and analyzed at step $n$ (one step per layer). At each step, $\Omega$ is partitioned into the following three subdomains (Fig.~\ref{fig:2-1}):
\begin{equation}\label{eq:2-3}
\Omega = \Omega_A^{(n)} \cup \Omega_{ihs}^{(n)} \cup \Omega_I^{(n)}
\end{equation}
where $\Omega_A^{(n)} = \bigcup_{k=1}^{n-1} L_k$ is the activated domain that has already been built, $\Omega_{ihs}^{(n)} = L_n$ is the layer that is activated and subjected to the inherent strain at the current step, and $\Omega_I^{(n)} = \bigcup_{k=n+1}^{N} L_k$ is the inactive domain. To describe them analytically, we introduce the indicator function of layer $L_n$,
\begin{equation}\label{eq:2-4}
\chi_{L_n}(\boldsymbol{x}) =
\begin{cases}
1 & \boldsymbol{x} \in L_n \\
0 & \text{otherwise}
\end{cases}
\end{equation}
and the activation indicator function that indicates the domain contributing to the stiffness at the end of step $n$,
\begin{equation}\label{eq:2-5}
\bar\chi^{(n)}(\boldsymbol{x}) = \sum_{k=1}^{n} \chi_{L_k}(\boldsymbol{x})
\end{equation}
with $\bar\chi^{(n)} = 1$ for $\boldsymbol{x} \in \Omega_A^{(n)} \cup \Omega_{ihs}^{(n)}$ and $\bar\chi^{(n)} = 0$ for $\boldsymbol{x} \in \Omega_I^{(n)}$.

For the numerical treatment of the inactive domain, two approaches are commonly used: the inactive element method and the quiet element method \cite{Lindgren1999}. The former excludes the inactive elements from the finite element assembly, whereas the latter keeps all the elements in the assembly and assigns very small material properties to the inactive elements. In this study, we adopt the quiet element method and assign a Young's modulus $E_0$, which is sufficiently smaller than that of the solid material, to the inactive domain $\Omega_I^{(n)}$, so that the layer activation is represented on a fixed mesh. The small but nonzero Young's modulus is introduced to avoid the rank deficiency of the global stiffness matrix that would arise if the stiffness of the inactive elements vanished while they are kept in the assembly; it is chosen sufficiently small that it does not affect the response of the activated domain.

Here, we describe the strategy of this study for modeling the layer activation. Previous studies using the quiet element method \cite{Takezawa2020,Xu2022,DugastTo2023} adopt formulations in which the displacement is accumulated over the entire building history; when a new layer is activated, the deformation accumulated up to that moment, $\boldsymbol{\varepsilon}(\boldsymbol{u}^{(n-1)})$, is inherited by the new layer, and a spurious initial stress arises. To cancel it, an activation strain $\boldsymbol{\varepsilon}^{act} = \boldsymbol{\varepsilon}(\boldsymbol{u}^{(n-1)})$ must be introduced into the constitutive equation so that the new layer is activated without this spurious initial stress. However, the activation strain carries the displacement field at the moment of element activation over into the constitutive equation, which introduces additional dependencies between layer steps; in elastoplastic analyses in particular, this complicates the sensitivity analysis and increases the computational cost \cite{DugastTo2023}. In contrast, in this study, the incremental displacement is solved anew at each step, and the stress history is carried over by explicitly incorporating the stress accumulated up to the previous step into the constitutive equation (Section~\ref{sec:2-3}). In this formulation, a new layer is activated with zero plastic history and zero stress history, so that no spurious initial stress arises in the new layer and no correction term such as the activation strain is needed; in the domain that has already been built, the accumulated stress is preserved so that the stress continuity across layer interfaces is guaranteed.

\subsection{Elastoplastic constitutive equations}\label{sec:2-3}

\subsubsection{Strain decomposition and accumulated form of the stress update}\label{sec:2-3-1}

Under the small strain assumption, the total strain at step $n$ is decomposed into the elastic strain, the plastic strain, and the inherent strain:
\begin{equation}\label{eq:2-6}
\boldsymbol{\varepsilon}^{(n)} = \boldsymbol{\varepsilon}^{e\,(n)} + \boldsymbol{\varepsilon}^{p\,(n)} + \boldsymbol{\varepsilon}^{*\,(n)}
\end{equation}
Here, the accumulated inherent strain is the sum of the inherent strains applied to the activated layers,
\begin{equation}\label{eq:2-7}
\boldsymbol{\varepsilon}^{*\,(n)}(\boldsymbol{x}) = \sum_{k=1}^{n} \chi_{L_k}(\boldsymbol{x}) \, \boldsymbol{\varepsilon}^*_0 = \bar\chi^{(n)}(\boldsymbol{x}) \, \boldsymbol{\varepsilon}^*_0
\end{equation}
Since each point is activated at most once, only one term of this sum is nonzero.

In this study, the incremental displacement $\boldsymbol{u}^{(n)}$, that is, the response to the inherent strain loading and the layer activation of the current step, is solved anew at each step. Using the symmetric gradient operator $\boldsymbol{\varepsilon}(\cdot)$, the total strain increment of step $n$ is written as $\Delta\boldsymbol{\varepsilon}^{(n)} := \boldsymbol{\varepsilon}(\boldsymbol{u}^{(n)})$, and the cumulative total strain is its sum over the steps, $\boldsymbol{\varepsilon}^{(n)} = \sum_{k=1}^{n} \Delta\boldsymbol{\varepsilon}^{(k)}$. Substituting the strain decomposition \eqref{eq:2-6} into the linear elastic relation between the elastic strain and the stress, $\boldsymbol{\sigma}^{(n)} = \mathbb{D}^e : \boldsymbol{\varepsilon}^{e\,(n)}$, then gives the stress accumulated up to step $n$; in the inactive domain, the elasticity tensor is the quiet tensor $\mathbb{D}^e_0$, and the inherent strain term is absent because no inherent strain has been applied there ($\bar\chi^{(n)} = 0$):
\begin{equation}\label{eq:2-7a}
\boldsymbol{\sigma}^{(n)} =
\begin{cases}
\mathbb{D}^e : \bigl[ \boldsymbol{\varepsilon}^{(n)} - \boldsymbol{\varepsilon}^{p\,(n)} - \boldsymbol{\varepsilon}^{*\,(n)} \bigr] & \text{in } \Omega_A^{(n)} \cup \Omega_{ihs}^{(n)} \\
\mathbb{D}^e_0 : \bigl[ \boldsymbol{\varepsilon}^{(n)} - \boldsymbol{\varepsilon}^{p\,(n)} \bigr] & \text{in } \Omega_I^{(n)}
\end{cases}
\end{equation}
where $\mathbb{D}^e = \kappa \boldsymbol{1} \otimes \boldsymbol{1} + 2\mu \mathbb{I}^{dev}$ is the elasticity tensor with bulk modulus $\kappa$ and shear modulus $\mu$, and $\mathbb{D}^e_0$ is the quiet elasticity tensor constructed from the small Young's modulus $E_0$ with the Poisson's ratio unchanged (Section~\ref{sec:2-2}). At the first step, all the history fields vanish, and the first branch of \eqref{eq:2-7a} reads $\boldsymbol{\sigma}^{(1)} = \mathbb{D}^e : [ \Delta\boldsymbol{\varepsilon}^{(1)} - \Delta\boldsymbol{\varepsilon}^{p\,(1)} - \chi_{L_1} \boldsymbol{\varepsilon}^*_0 ]$; at the second step, subtracting the expression of $\boldsymbol{\sigma}^{(1)}$ from \eqref{eq:2-7a} with $n = 2$ leaves only the quantities of step $2$ in the bracket. In general, taking the difference of each branch of \eqref{eq:2-7a} between two consecutive steps and using \eqref{eq:2-7}, the stress update closes in the following incremental form:
\begin{equation}\label{eq:2-8}
\boldsymbol{\sigma}^{(n)} =
\begin{cases}
\mathbb{D}^e : \bigl[ \Delta\boldsymbol{\varepsilon}^{(n)} - \Delta\boldsymbol{\varepsilon}^{p\,(n)} - \chi_{L_n} \boldsymbol{\varepsilon}^*_0 \bigr] + \boldsymbol{\sigma}^{(n-1)} & \text{in } \Omega_A^{(n)} \cup \Omega_{ihs}^{(n)} \\
\mathbb{D}^e_0 : \bigl[ \Delta\boldsymbol{\varepsilon}^{(n)} - \Delta\boldsymbol{\varepsilon}^{p\,(n)} \bigr] + \boldsymbol{\sigma}^{(n-1)} & \text{in } \Omega_I^{(n)}
\end{cases}
\end{equation}
where $\Delta\boldsymbol{\varepsilon}^{p\,(n)} = \boldsymbol{\varepsilon}^{p\,(n)} - \boldsymbol{\varepsilon}^{p\,(n-1)}$ is the plastic strain increment, and $\chi_{L_n} \boldsymbol{\varepsilon}^*_0 = \boldsymbol{\varepsilon}^{*\,(n)} - \boldsymbol{\varepsilon}^{*\,(n-1)}$ is the inherent strain increment newly applied at the current step. The accumulated stress $\boldsymbol{\sigma}^{(n-1)}$ denotes the stress accumulated up to the previous step, defined recursively by \eqref{eq:2-8} itself from the initial state $\boldsymbol{\sigma}^{(0)} = \boldsymbol{0}$. Note that the plastic strains generated in the preceding steps do not appear explicitly in \eqref{eq:2-8}: as is evident from the accumulated form \eqref{eq:2-7a}, they are already contained in the accumulated stress $\boldsymbol{\sigma}^{(n-1)}$, and only the increment $\Delta\boldsymbol{\varepsilon}^{p\,(n)}$ of the current step enters the update.

Equation~\eqref{eq:2-8} is the core of the present analysis model. The stress history up to the previous step is explicitly incorporated into the constitutive equation as the accumulated stress $\boldsymbol{\sigma}^{(n-1)}$ and acts as an initial stress in the activated domain. On the other hand, the new layer $\Omega_{ihs}^{(n)} = L_n$ carries no history of the preceding steps, as described in Section~\ref{sec:2-2}, and hence the accumulated stress vanishes there, $\boldsymbol{\sigma}^{(n-1)} = \boldsymbol{0}$; the new layer is thus activated without any spurious initial stress.

\begin{figure}[pos=!htbp]
  \centering
  \includegraphics[width=\linewidth]{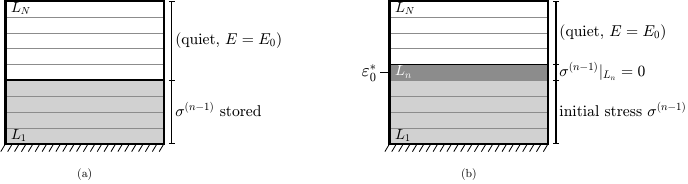}
  \caption{Incremental structure of the stress update at layer step $n$. (a) At the end of step $n-1$, the built layers store the accumulated stress $\boldsymbol{\sigma}^{(n-1)}$, and the layers above remain quiet. (b) At step $n$, the layer $L_n$ is activated with zero stress history, $\boldsymbol{\sigma}^{(n-1)}|_{L_n} = \boldsymbol{0}$, and subjected to the inherent strain $\boldsymbol{\varepsilon}^*_0$, while the accumulated stress $\boldsymbol{\sigma}^{(n-1)}$ acts as an initial stress in the built layers; the incremental displacement $\boldsymbol{u}^{(n)}$ is solved anew on the entire domain, and the stress is updated by \eqref{eq:2-8}.}
  \label{fig:2-1a}
\end{figure}

Although the plastic evolution is formally retained in the second branch of \eqref{eq:2-8}, the stress level in the quiet state is smaller than that of the built material by the factor $E_0/E$, where $E$ denotes the Young's modulus of the solid material, and remains far inside the yield surface, so that $\Delta\boldsymbol{\varepsilon}^{p\,(n)} = \boldsymbol{0}$ there in practice. The quiet stress accumulated by the second branch of \eqref{eq:2-8} is an artifact of the numerical realization of the layer activation; as described in Section~\ref{sec:2-4}, it is discarded at the moment of activation and never enters the history of the built material. The incremental structure of the update at step $n$, that is, the carry-over of $\boldsymbol{\sigma}^{(n-1)}$ as an initial stress in the built layers and the activation of the new layer with zero stress history, is illustrated in Fig.~\ref{fig:2-1a}.

\subsubsection{Yield function and linear isotropic hardening}\label{sec:2-3-2}

The plastic behavior is modeled by $J_2$ plasticity with the von Mises yield function and the associated flow rule. The yield function is given by
\begin{equation}\label{eq:2-9}
f\bigl(\boldsymbol{\sigma}, \bar\alpha\bigr) = \sigma_{vm}(\boldsymbol{\sigma}) - \sigma_y(\bar\alpha), \quad \sigma_y(\bar\alpha) := \sigma_{y0} + H \bar\alpha
\end{equation}
where $\sigma_{vm}(\boldsymbol{\sigma}) = \sqrt{\tfrac{3}{2}}\,\|\boldsymbol{s}\|$ is the von Mises equivalent stress of the deviatoric stress $\boldsymbol{s} = \mathrm{dev}(\boldsymbol{\sigma})$, $\bar\alpha$ is the equivalent plastic strain, $\sigma_{y0}$ is the initial yield stress, and $H$ is the linear isotropic hardening modulus.

The plastic evolution follows the associated flow rule, and the loading\slash unloading at each step is governed by the incremental Karush--Kuhn--Tucker (KKT) conditions \cite{SimoHughes1998,Granlund2024}:
\begin{equation}\label{eq:2-10}
\Delta\bar\alpha^{(n)} \geq 0, \qquad
f\bigl(\boldsymbol{\sigma}^{(n)}, \bar\alpha^{(n)}\bigr) \leq 0, \qquad
\Delta\bar\alpha^{(n)}\, f\bigl(\boldsymbol{\sigma}^{(n)}, \bar\alpha^{(n)}\bigr) = 0
\end{equation}
That is, in the elastic state ($\Delta\bar\alpha^{(n)} = 0$) the state remains inside the yield surface, and in the plastic state ($\Delta\bar\alpha^{(n)} > 0$) the updated state is required to lie on the yield surface, $f = 0$; this is the consistency condition.

\subsubsection{Closed-form update by radial return and the recursive structure of the state}\label{sec:2-3-3}

The stress update at each step, that is, the determination of the state $(\boldsymbol{\sigma}^{(n)}, \bar\alpha^{(n)})$ satisfying the KKT conditions \eqref{eq:2-10}, can be written in closed form by the standard predictor--corrector scheme, that is, the radial return \cite{SimoHughes1998}. First, the elastic trial stress obtained by assuming a zero plastic increment at the current step is defined as
\begin{equation}\label{eq:2-11}
\boldsymbol{\sigma}^{tr\,(n)} = \mathbb{D}^e : \bigl[ \Delta\boldsymbol{\varepsilon}^{(n)} - \chi_{L_n} \boldsymbol{\varepsilon}^*_0 \bigr] + \boldsymbol{\sigma}^{(n-1)}
\end{equation}
For linear isotropic hardening, the KKT conditions \eqref{eq:2-10} are solved in closed form, and the equivalent plastic strain increment is given by
\begin{equation}\label{eq:2-12}
\Delta\bar\alpha^{(n)} = \frac{\bigl\langle \sigma_{vm}\bigl(\boldsymbol{\sigma}^{tr\,(n)}\bigr) - \sigma_y\bigl(\bar\alpha^{(n-1)}\bigr) \bigr\rangle}{3\mu + H}, \qquad \bar\alpha^{(n)} = \bar\alpha^{(n-1)} + \Delta\bar\alpha^{(n)}
\end{equation}
where the Macaulay bracket $\langle\cdot\rangle$ appears as the closed-form solution of the complementarity condition in \eqref{eq:2-10} and represents the branch between the elastic ($\Delta\bar\alpha^{(n)} = 0$) and plastic ($\Delta\bar\alpha^{(n)} > 0$) states. The updated stress is written as a radial scaling of the trial deviatoric stress:
\begin{equation}\label{eq:2-13}
\begin{aligned}
\boldsymbol{\sigma}^{(n)} &= \mathrm{vol}\bigl(\boldsymbol{\sigma}^{tr\,(n)}\bigr) + \theta^{(n)} \boldsymbol{s}^{tr\,(n)}, \\
\theta^{(n)} &= \frac{\sigma_y\bigl(\bar\alpha^{(n)}\bigr)}{\sigma_{vm}\bigl(\boldsymbol{\sigma}^{tr\,(n)}\bigr)} = 1 - \frac{3\mu\,\Delta\bar\alpha^{(n)}}{\sigma_{vm}\bigl(\boldsymbol{\sigma}^{tr\,(n)}\bigr)}
\end{aligned}
\end{equation}
In the elastic state ($f \leq 0$), $\Delta\bar\alpha^{(n)} = 0$ and $\theta^{(n)} = 1$, and \eqref{eq:2-13} reduces to $\boldsymbol{\sigma}^{(n)} = \boldsymbol{\sigma}^{tr\,(n)}$. The plastic strain increment follows from the associated flow rule as $\Delta\boldsymbol{\varepsilon}^{p\,(n)} = \sqrt{\tfrac{3}{2}}\,\Delta\bar\alpha^{(n)}\, \hat{\boldsymbol{n}}^{(n)}$ with $\hat{\boldsymbol{n}}^{(n)} = \boldsymbol{s}^{tr\,(n)}/\|\boldsymbol{s}^{tr\,(n)}\|$, and substituting it into \eqref{eq:2-8} recovers \eqref{eq:2-13}.

Taking the pair of the accumulated stress and the equivalent plastic strain $(\boldsymbol{\sigma}^{(n)}, \bar\alpha^{(n)})$ as the state at each material point (at each Gauss point in the finite element implementation), the above update rule closes as a one-step recurrence, that is, an update mapping that depends only on the state of the previous step and the displacement of the current step:
\begin{equation}\label{eq:2-14}
\bar\alpha^{(n)} = A^{(n)}\bigl(\boldsymbol{\sigma}^{tr\,(n)}, \bar\alpha^{(n-1)}\bigr), \qquad
\boldsymbol{\sigma}^{(n)} = \boldsymbol{S}^{(n)}\bigl(\boldsymbol{\sigma}^{tr\,(n)}, \bar\alpha^{(n-1)}\bigr)
\end{equation}
Here, the scalar update mapping $A^{(n)}$ denotes the closed-form update \eqref{eq:2-12} of the equivalent plastic strain, and the tensor update mapping $\boldsymbol{S}^{(n)}$ denotes the radial return \eqref{eq:2-13} combined with the trial stress \eqref{eq:2-11}. Since the plastic strain $\boldsymbol{\varepsilon}^{p\,(n)}$ is absorbed into the accumulated stress $\boldsymbol{\sigma}^{(n)}$, it need not be retained as an independent state variable. This choice of the state and the recursive structure form the basis of the adjoint sensitivity analysis derived in Section~\ref{sec:4}. Moreover, since the elastic/plastic branch is determined by whether $\Delta\bar\alpha^{(n)} > 0$ holds in the KKT conditions \eqref{eq:2-10} of the forward analysis, this active set directly provides the gate of the plastic terms in the adjoint equations in Section~\ref{sec:4}.

Furthermore, the derivatives of the update mappings \eqref{eq:2-13} and \eqref{eq:2-14} are also obtained in closed form. The derivative of the stress update mapping with respect to the trial stress is given by the return-mapping projection tensor
\begin{equation}\label{eq:2-15}
\mathbb{T}^{(n)} := \frac{\partial \boldsymbol{S}^{(n)}}{\partial \boldsymbol{\sigma}^{tr}}
= \mathbb{I}^{vol} + \theta^{(n)}\, \mathbb{I}^{dev} - \bar\theta^{(n)}\, \hat{\boldsymbol{n}}^{(n)} \otimes \hat{\boldsymbol{n}}^{(n)},
\qquad
\bar\theta^{(n)} := \bigl(\theta^{(n)} - 1\bigr) + \frac{3\mu}{3\mu + H}
\end{equation}
with $\mathbb{T}^{(n)} = \mathbb{I}$ in the elastic state. Using this tensor, the algorithmic tangent stiffness, also called the consistent tangent, of the stress with respect to the strain increment at step $n$ is expressed as
\begin{equation}\label{eq:2-16}
\mathbb{D}^{ats\,(n)} := \frac{\partial \boldsymbol{\sigma}^{(n)}}{\partial \Delta\boldsymbol{\varepsilon}^{(n)}}
= \mathbb{T}^{(n)} : \mathbb{D}^e
= \kappa \boldsymbol{1} \otimes \boldsymbol{1} + 2\mu\, \theta^{(n)} \mathbb{I}^{dev} - 2\mu\, \bar\theta^{(n)} \hat{\boldsymbol{n}}^{(n)} \otimes \hat{\boldsymbol{n}}^{(n)}
\end{equation}
$\mathbb{D}^{ats\,(n)}$ is used in the Newton iteration of the equilibrium equation (Section~\ref{sec:2-4}). We emphasize that, as shown in Section~\ref{sec:4}, the same coefficient tensors $\mathbb{T}^{(n)}$ and $\mathbb{D}^{ats\,(n)}$ are reused as they are in the reverse update of the adjoint fields. In the elastic limit, $\mathbb{D}^{ats\,(n)} = \mathbb{D}^e$, which coincides with the stiffness of the linear elastic AM model.

\subsection{Governing equations and the layer-by-layer analysis algorithm}\label{sec:2-4}

The bottom surface $\Gamma_u$ of the part is clamped to the build plate, and the other boundaries are traction-free. The admissible displacement space with the displacement constrained on $\Gamma_u$ is defined as
\begin{equation}\label{eq:2-17}
V := \bigl\{ \boldsymbol{v} \in \bigl[H^1(\Omega)\bigr]^d \; : \; \boldsymbol{v} = \boldsymbol{0} \ \text{on} \ \Gamma_u \bigr\}
\end{equation}
The displacement field $\boldsymbol{u}^{(n)} \in V$ at step $n$ is governed by the weak form of the quasi-static equilibrium without external loads:
\begin{equation}\label{eq:2-18}
\int_\Omega \bar\chi^{(n)} \, \boldsymbol{\sigma}^{(n)}\bigl(\boldsymbol{u}^{(n)}\bigr) : \boldsymbol{\varepsilon}(\boldsymbol{v}) \, dV = 0
\qquad \forall \boldsymbol{v} \in V
\end{equation}
where $\boldsymbol{v} \in V$ is an arbitrary test function, that is, a virtual displacement, and $\boldsymbol{\varepsilon}(\boldsymbol{v})$ denotes its symmetric gradient. The above equation is solved sequentially for all indices $n = 1, 2, \ldots, N$. No external load acts during the building process, and the stress is generated only by the application of the inherent strain. Since $\boldsymbol{\sigma}^{(n)}$ is a nonlinear mapping of $\boldsymbol{\varepsilon}(\boldsymbol{u}^{(n)})$ that includes the return mapping \eqref{eq:2-11}--\eqref{eq:2-13}, Equation~\eqref{eq:2-18} is solved by the Newton method with the algorithmic tangent stiffness $\mathbb{D}^{ats\,(n)}$ in \eqref{eq:2-16}.

Equation~\eqref{eq:2-18} is the sharp form in which the inactive domain is excluded by the activation indicator $\bar\chi^{(n)}$. In the finite element computation, all the elements are kept in the assembly by the quiet element method (Section~\ref{sec:2-2}), and the equilibrium actually solved is assembled over the entire fixed domain:
\begin{equation}\label{eq:2-18a}
\int_\Omega \boldsymbol{\sigma}^{(n)}\bigl(\boldsymbol{u}^{(n)}\bigr) : \boldsymbol{\varepsilon}(\boldsymbol{v}) \, dV = 0
\qquad \forall \boldsymbol{v} \in V
\end{equation}
where $\boldsymbol{\sigma}^{(n)}$ is given by the two branches of \eqref{eq:2-8}. Since the contribution of $\Omega_I^{(n)}$ is of the order of $E_0/E$, Equation~\eqref{eq:2-18a} reduces to \eqref{eq:2-18} in the limit $E_0 \to 0$; the formulation and the sensitivity analysis in the subsequent sections are developed on the basis of the sharp form \eqref{eq:2-18}.

The initial conditions are
\begin{equation}\label{eq:2-19}
\boldsymbol{\sigma}^{(0)} = \boldsymbol{0}, \qquad \bar\alpha^{(0)} = 0
\end{equation}
and, at the activation of a new layer $L_n$, the state history within the layer is initialized as
\begin{equation}\label{eq:2-20}
\bigl(\boldsymbol{\sigma}^{(n-1)}, \bar\alpha^{(n-1)}\bigr)\bigl|_{L_n} = (\boldsymbol{0}, 0)
\end{equation}
(this initialization discards the quiet stress accumulated by the second branch of \eqref{eq:2-8}, which is negligible as noted in Section~\ref{sec:2-3-1}). The layer-by-layer analysis procedure is summarized in Algorithm~\ref{alg:1}.

\begin{algorithm}[tb]
\caption{Layer-by-layer analysis based on the elastoplastic inherent strain method}
\label{alg:1}
\begin{algorithmic}[1]
\State Initialize: $\boldsymbol{\sigma}^{(0)} = \boldsymbol{0}$, $\bar\alpha^{(0)} = 0$
\For{$n = 1, 2, \ldots, N$}
  \State Activate layer $L_n$: keep the Young's modulus of $\Omega_I^{(n)}$ at $E_0$ and assign the solid material properties to $L_n$
  \State Initialize the state history within layer $L_n$: $(\boldsymbol{\sigma}^{(n-1)}, \bar\alpha^{(n-1)})|_{L_n} = (\boldsymbol{0}, 0)$
  \State Apply the inherent strain $\chi_{L_n} \boldsymbol{\varepsilon}^*_0$ and solve the equilibrium equation \eqref{eq:2-18} by Newton iterations to obtain $\boldsymbol{u}^{(n)}$, executing the return mapping \eqref{eq:2-11}--\eqref{eq:2-13} at each Gauss point
  \State Update and store the state: $(\boldsymbol{\sigma}^{(n)}, \bar\alpha^{(n)})$
\EndFor
\State Output: final residual stress field $\boldsymbol{\sigma}^{(N)}$ and equivalent plastic strain field $\bar\alpha^{(N)}$
\end{algorithmic}
\end{algorithm}

The computation at each step of Algorithm~\ref{alg:1} is a static elastoplastic analysis whose only input is the state of the previous step, $(\boldsymbol{\sigma}^{(n-1)}, \bar\alpha^{(n-1)})$, and the total analysis cost of the $N$ steps is bounded by $N$ times that of a single elastoplastic analysis. The final field $\boldsymbol{\sigma}^{(N)}$ gives the residual stress at the completion of the building process and is used to evaluate the objective function of the optimization problem formulated in Section~\ref{sec:3}.

\subsection{Numerical example}\label{sec:2-5}

In this section, we confirm the behavior of the proposed elastoplastic layer-by-layer analysis model and demonstrate the necessity of considering plasticity through a comparison with a linear elastic analysis model.

\subsubsection{Problem settings}\label{sec:2-5-1}

\begin{figure}[pos=!htbp]
  \centering
  \includegraphics[width=0.60\linewidth]{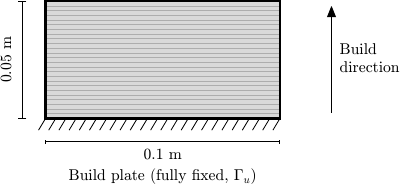}
  \caption{Analysis model of Section~\ref{sec:2-5}. The fully solid $100 \times 50$ mm rectangular domain (plane stress) is divided into $N = 25$ layers of thickness $h_L = 2$ mm. The bottom surface $\Gamma_u$ is fixed to the build plate, and the inherent strain $\boldsymbol{\varepsilon}^*_0$ is applied to the newly activated layer at each step (no external load).}
  \label{fig:2-2}
\end{figure}

The analysis domain is the two-dimensional rectangular domain of width $100\,\mathrm{mm}$ and height $50\,\mathrm{mm}$ shown in Fig.~\ref{fig:2-2}, which is entirely filled with the solid material and analyzed under plane stress. The bottom surface $\Gamma_u$ of the part is fixed to the build plate. The domain is divided into $N = 25$ layers of thickness $h_L = 2\,\mathrm{mm}$ along the building direction, and the layer-by-layer analysis is performed by Algorithm~\ref{alg:1}. An aluminum alloy is assumed as the material, with Young's modulus $E = 72\,\mathrm{GPa}$, Poisson's ratio $\nu = 0.3$, initial yield stress $\sigma_{y0} = 243\,\mathrm{MPa}$, and linear hardening modulus $H = 2.171\,\mathrm{GPa}$. Following the common practice \cite{Xu2022}, the inherent strain tensor is prescribed as the uniaxial constant tensor $\boldsymbol{\varepsilon}^*_0 = (\varepsilon^*_1, 0, 0)$ in Voigt notation, in which only the in-plane normal component $\varepsilon^*_1$ is nonzero and the building-direction and shear components are omitted. In this example, $\varepsilon^*_1 = -3.0 \times 10^{-3}$. The analysis conditions are summarized in Table~\ref{tab:2-1}.

\begin{table}[pos=!htbp]
  \centering
  \small
  \caption{Material and process parameters used in Section~\ref{sec:2-5}}
  \label{tab:2-1}
  \begin{tabularx}{\textwidth}{>{\raggedright\arraybackslash}p{2.5cm} >{\raggedright\arraybackslash}X >{\raggedright\arraybackslash}p{3.0cm}}
    \toprule
    Category & Symbol (definition) & Value \\
    \midrule
    Material & Young's modulus $E$ & $72$ GPa \\
             & Poisson's ratio $\nu$ & $0.3$ \\
             & Initial yield stress $\sigma_{y0}$ \eqref{eq:2-9} & $243$ MPa \\
             & Linear hardening modulus $H$ \eqref{eq:2-9} & $2.171$ GPa \\
    Process  & Inherent strain $\varepsilon^*_1$ & $-3.0 \times 10^{-3}$ \\
             & Layer thickness $h_L$ / number of layers $N$ & $2$ mm / $25$ \\
             & Young's modulus of quiet elements $E_0$ (Section~\ref{sec:2-2}) & $1$ kPa \\
    \bottomrule
  \end{tabularx}
\end{table}

The domain is discretized with a $50 \times 25$ structured grid of $2 \times 2$ mm elements, and each layer corresponds to one row of elements. Quadrilateral elements with quadratic interpolation are used for the displacement, and the stress and the plastic state variables are stored at the $3 \times 3$ Gauss points of each element.

As the comparison target, we employ the linear elastic AM model in which the plasticity is disabled under the identical layer division, quiet element treatment, and inherent strain loading, that is, the accumulated-form elastic analysis obtained by setting $\Delta\boldsymbol{\varepsilon}^{p\,(n)} = \boldsymbol{0}$ in \eqref{eq:2-8} \cite{Xu2022}. Since the two models differ only in the constitutive law, their comparison directly shows the influence of plastic deformation on the residual stress prediction.

\subsubsection{Results}\label{sec:2-5-2}

\begin{figure}[pos=!htbp]
  \centering
  \begin{subfigure}[t]{0.49\linewidth}
    \centering
    \includegraphics[width=\linewidth]{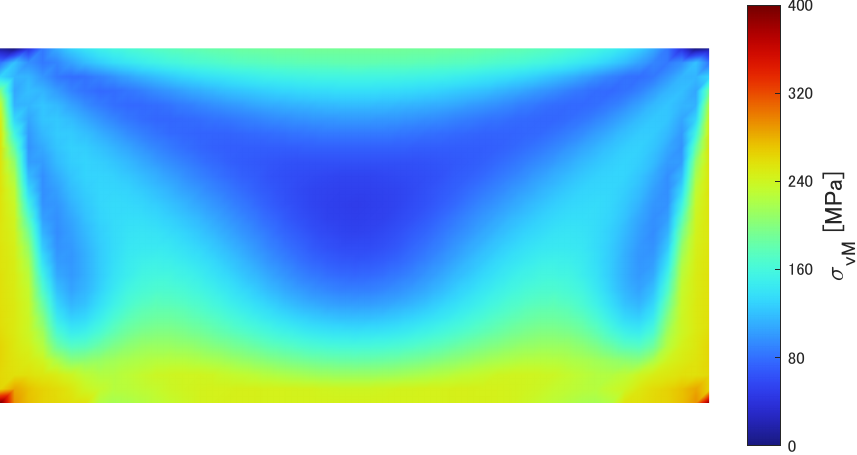}
    \caption{}
    \label{fig:2-3a}
  \end{subfigure}%
  \hfill%
  \begin{subfigure}[t]{0.49\linewidth}
    \centering
    \includegraphics[width=\linewidth]{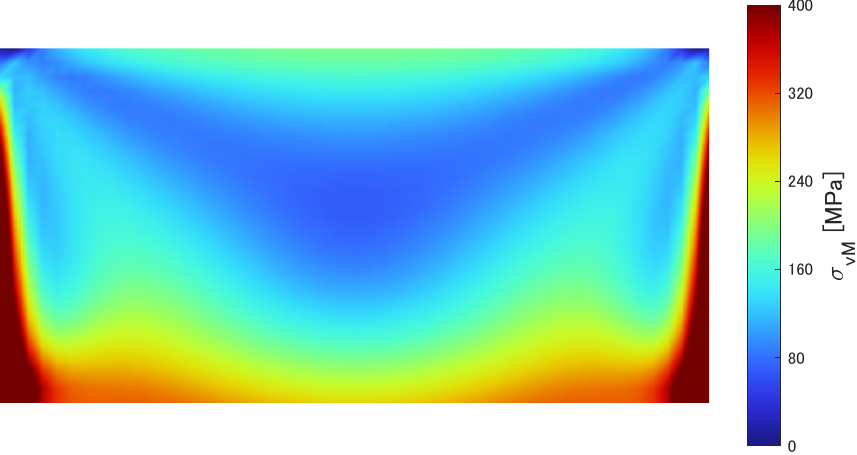}
    \caption{}
    \label{fig:2-3b}
  \end{subfigure}
  \caption{Distributions of the von Mises residual stress at the completion of the building process. (a) Elastoplastic model. (b) Linear elastic model. The color range 0--400 MPa is common to both (values above the upper limit are saturated). The maximum value of the linear elastic model, $1730.9\,\mathrm{MPa}$, greatly exceeds this range.}
  \label{fig:2-3}
\end{figure}

\begin{figure}[pos=!htbp]
  \centering
  \includegraphics[width=0.62\linewidth]{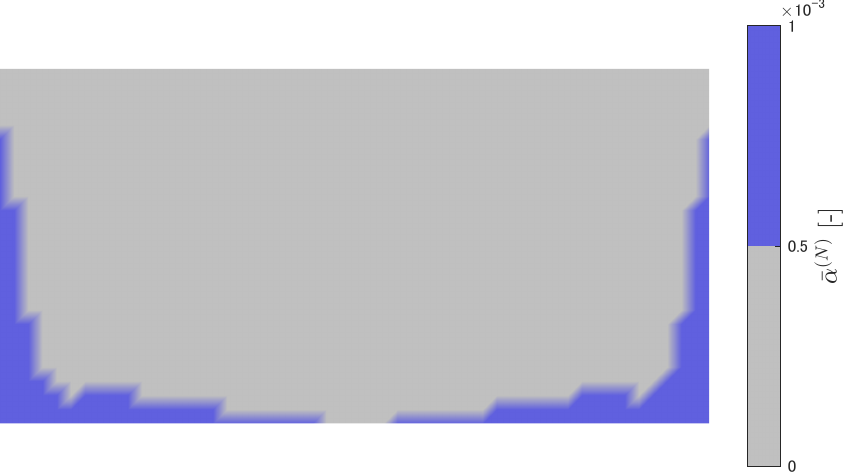}
  \caption{Distribution of the equivalent plastic strain $\bar\alpha^{(N)}$ at the completion of the building process in the elastoplastic model (blue: $\bar\alpha^{(N)} \geq 5 \times 10^{-4}$).}
  \label{fig:2-4}
\end{figure}

The distribution of the von Mises residual stress $\sigma_{vm}(\boldsymbol{\sigma}^{(N)})$ at the completion of the building process is shown in Fig.~\ref{fig:2-3}, and that of the equivalent plastic strain $\bar\alpha^{(N)}$ in the elastoplastic model is shown in Fig.~\ref{fig:2-4}.

In the elastoplastic model (Fig.~\ref{fig:2-3}(a)), the maximum stress of $380.0\,\mathrm{MPa}$ occurs at the junction with the build plate and remains $1.56$ times the initial yield stress. Since the stress cannot exceed the yield surface $\sigma_y(\bar\alpha) = \sigma_{y0} + H\bar\alpha$, this excess corresponds to the expansion of the yield surface by linear isotropic hardening. Indeed, substituting the maximum equivalent plastic strain $\bar\alpha^{(N)}_{\max} = 6.31 \times 10^{-2}$ into the hardening law yields $\sigma_y = 380.0\,\mathrm{MPa}$, which agrees with the maximum stress. Outside the bottom layer, the highly stressed regions are limited to bands near the side surfaces, and the stress level there is bounded slightly above the yield stress ($250$--$275\,\mathrm{MPa}$).

The distribution of the equivalent plastic strain (Fig.~\ref{fig:2-4}) shows that plastic deformation occurs in the bottom layer adjacent to the build plate and in bands rising along both side surfaces, and little plastic strain remains in the interior. The region where the plastic strain remains at the completion of the building process occupies 12.3\% of the whole domain. Layer-wise statistics show that 97\% of the Gauss points undergo plastic deformation in the bottom layer, that the ratio decreases monotonically toward the top, and that no plastic deformation occurs in the four topmost layers, that is, above the height of $42\,\mathrm{mm}$. This is because the layers built earlier are repeatedly constrained by the shrinkage of the newly activated layers, whereas the layer activated last has no constraint above it; this reflects the history dependence peculiar to the layer-by-layer analysis.

In contrast, in the linear elastic model (Fig.~\ref{fig:2-3}(b)), the maximum value reaches $1730.9\,\mathrm{MPa}$, which is $4.6$ times that of the elastoplastic model. Meanwhile, the domain averages are $170.6\,\mathrm{MPa}$ versus $141.3\,\mathrm{MPa}$, a difference of only 21\%. In other words, the overestimation by the linear elastic model concentrates in the highly stressed regions, and the error is larger precisely where the absolute stress value matters. In the regions beyond the yield stress, the stress in the linear elastic model is not bounded by the yield surface and no redistribution by plastic deformation occurs, so the stress predicted there has no physical meaning in either magnitude or distribution.

The present comparison quantitatively confirms, under the analysis model and building conditions of this study, the overestimation by the linear elastic inherent strain method reported in Section~\ref{sec:2-1} \cite{Bellet2023}. In topology optimization that uses the residual stress as the objective function or as a constraint, the stress values themselves drive the design, and hence this overestimation directly impairs the validity of the optimized design. Therefore, the adoption of an analysis model based on the elastoplastic constitutive law is indispensable for the optimization method constructed in the following sections.

In the next section, the present analysis model is incorporated into the framework of density-based topology optimization, and the optimization problem for reducing the residual stress during the building process is formulated.

\section{Formulation of the topology optimization problem}\label{sec:3}

In this section, the elastoplastic layer-by-layer analysis model formulated in Section~\ref{sec:2} is incorporated into the framework of density-based topology optimization, and an optimization problem that minimizes the residual stress under a compliance constraint is formulated. Section~\ref{sec:3-1} introduces the design variable, the filter and projection, and the material interpolation, and makes the analysis model design-dependent. Section~\ref{sec:3-2} formulates the minimum mean compliance problem as a reference problem, and Section~\ref{sec:3-3} formulates the residual stress minimization problem. Section~\ref{sec:3-4} presents the residual stress minimization problem under a compliance constraint, which integrates the two problems.

\subsection{Design variable and material interpolation}\label{sec:3-1}

In density-based topology optimization, the structural geometry is represented by a density field on a fixed design domain \cite{Bendsoe1989,BendsoeSigmund2003}. In this and the following sections, the fixed domain $\Omega$ analyzed in Section~\ref{sec:2} is directly used as the fixed design domain. The material domain of the structure is represented as the region with the physical density $\bar z \approx 1$ in $\Omega$, and the void region with $\bar z \approx 0$ is included in the analysis as a weak material with the residual stiffness described below. This is the so-called ersatz material approximation. Treating the void as a weak material is of the same type as the quiet element treatment of the inactive domain in Section~\ref{sec:2-2}, and hence the layer-by-layer analysis model of Section~\ref{sec:2} is applicable to the whole of $\Omega$ without changing the governing equations or the layer division $\{L_n\}$. To the design variable $\psi(\boldsymbol{x}) \in [0, 1]$, a density filter of radius $r$ \cite{Bourdin2001,BrunsTortorelli2001} is applied in order to avoid numerical instabilities such as checkerboard patterns and to control the minimum member size:
\begin{equation}\label{eq:3-1}
\tilde\rho(\boldsymbol{x}) = \frac{\displaystyle\int_\Omega w(\boldsymbol{x}, \boldsymbol{y})\, \psi(\boldsymbol{y})\, dV_y}{\displaystyle\int_\Omega w(\boldsymbol{x}, \boldsymbol{y})\, dV_y},
\qquad
w(\boldsymbol{x}, \boldsymbol{y}) = \max\bigl(0,\; r - \|\boldsymbol{x} - \boldsymbol{y}\|\bigr)
\end{equation}
Since the filtered density $\tilde\rho$ contains intermediate densities, it is subsequently transformed into the physical density field $\bar z$ by the Heaviside projection \cite{Guest2004,Wang2011}, which promotes clear material/void boundaries:
\begin{equation}\label{eq:3-2}
\bar z = H_{\beta,\eta}(\tilde\rho) = \frac{\tanh(\beta\eta) + \tanh\bigl(\beta(\tilde\rho - \eta)\bigr)}{\tanh(\beta\eta) + \tanh\bigl(\beta(1 - \eta)\bigr)}
\end{equation}
where $\eta$ is the threshold and $\beta$ is the parameter that controls the sharpness of the projection.

The material properties are made dependent on the physical density $\bar z$. The RAMP interpolation \cite{StolpeSvanberg2001} is used for the Young's modulus:
\begin{equation}\label{eq:3-3}
E(\bar z) = \eta_E(\bar z)\, E_1, \qquad
\eta_E(\bar z) = \epsilon_E + (1 - \epsilon_E)\, \frac{\bar z}{1 + q_E (1 - \bar z)}
\end{equation}
where $E_1$ is the Young's modulus of the solid material, $q_E$ is the penalization parameter, and $\epsilon_E$ is a residual value for numerical stabilization. Note that this interpolation applies to the activated domain; in the inactive domain, the Young's modulus is kept at the small value $E_0$ by the quiet element method independently of $\bar z$ (Section~\ref{sec:2-2}, Eq.~\eqref{eq:2-8}). The reason for adopting the RAMP interpolation is as follows. Since the inherent strain acts as a design-dependent load through the stiffness, as described below, a power law such as SIMP applied to the stiffness causes both the stiffness and the inherent strain load to degenerate simultaneously in the low-density region, so that the sensitivity vanishes and intermediate densities become difficult to eliminate \cite{Xu2022}. The RAMP interpolation retains a finite gradient as $\bar z \to 0$ and avoids this problem.

On the other hand, a power-law interpolation is used for the initial yield stress and the linear hardening modulus:
\begin{equation}\label{eq:3-4}
\sigma_{y0}(\bar z) = \eta_p(\bar z)\, \sigma_{y1}, \qquad
H(\bar z) = \eta_p(\bar z)\, H_1, \qquad
\eta_p(\bar z) = \epsilon_p + (1 - \epsilon_p)\, \bar z^{\, q_p}
\end{equation}
where $\sigma_{y1}$ and $H_1$ are the values of the solid material and $q_p$ is the exponent. We take $q_p < 1$; specifically, the same value as the relaxation exponent $q_s$ of the stress evaluation (Section~\ref{sec:3-3}) is used. This is because an interpolation in which the yield stress degenerates faster than the stiffness ($q_p > 1$) causes the trial stress to exceed the interpolated yield stress in the low-density region, leading to nonphysical plastic deformation. The residual value $\epsilon_p$ is likewise introduced to prevent spurious plastic responses in the void region where the stiffness has almost vanished. It should be noted that the optimization result may depend on the choice of the exponent $q_p$, as is generally the case for parameter interpolation schemes. The recently proposed extended discrete material optimization (XDMO) framework \cite{Han2026}, which interpolates the governing equations of the constituent materials rather than their material parameters, could remove this dependence on $q_p$; however, its introduction is beyond the scope of this study.

The inherent strain $\boldsymbol{\varepsilon}^*_0$ itself is not made design-dependent. In the constitutive equation of Section~\ref{sec:2}, the inherent strain contributes to the stress in the form $\mathbb{D}^e : \chi_{L_n}\boldsymbol{\varepsilon}^*_0$; the equivalent inherent strain load therefore becomes density-dependent automatically through the stiffness interpolation \eqref{eq:3-3}, and the load vanishes together with the stiffness in the void region.

With the above interpolations, the analysis model of Section~\ref{sec:2} becomes design-dependent. That is, the elasticity tensor becomes
\begin{equation}\label{eq:3-5}
\mathbb{D}^e(\bar z) = \kappa(\bar z)\, \boldsymbol{1} \otimes \boldsymbol{1} + 2\mu(\bar z)\, \mathbb{I}^{dev}
\end{equation}
where $\kappa(\bar z)$ and $\mu(\bar z)$ are determined from $E(\bar z)$ and the Poisson's ratio $\nu$, and the layer-by-layer analysis for a design $\psi$ is obtained by replacing $\mathbb{D}^e$, $\sigma_{y0}$, and $H$ in \eqref{eq:2-8}--\eqref{eq:2-16} with the interpolated values of \eqref{eq:3-5} and \eqref{eq:3-4}.

\subsection{Minimum mean compliance problem}\label{sec:3-2}

First, the standard minimum mean compliance problem is presented as a reference problem. Considering the final use of the part after the completion of the building process, the displacement is constrained on the support boundary $\Gamma_D$, and a traction $\boldsymbol{t}$ acts on the boundary $\Gamma_t$. Note that the support condition $\Gamma_D$ in the final use generally differs from the build plate constraint $\Gamma_u$ in the building process (Section~\ref{sec:2-4}); to distinguish them from the displacement field $\boldsymbol{u}^{(n)}$ and the test function $\boldsymbol{v}$ of the building process, the displacement field and the test function of the final use are denoted with a hat. The final-use displacement field $\hat{\boldsymbol{u}} \in U := \{\hat{\boldsymbol{v}} \in [H^1(\Omega)]^d : \hat{\boldsymbol{v}} = \boldsymbol{0} \ \text{on} \ \Gamma_D\}$ is governed by the weak form of static linear elasticity:
\begin{equation}\label{eq:3-6}
\int_\Omega \mathbb{D}^e(\bar z) : \boldsymbol{\varepsilon}(\hat{\boldsymbol{u}}) : \boldsymbol{\varepsilon}(\hat{\boldsymbol{v}})\, dV
= \int_{\Gamma_t} \boldsymbol{t} \cdot \hat{\boldsymbol{v}}\, d\Gamma
\qquad \forall \hat{\boldsymbol{v}} \in U
\end{equation}
The structural stiffness is evaluated by the mean compliance
\begin{equation}\label{eq:3-7}
l(\hat{\boldsymbol{u}}) = \int_{\Gamma_t} \boldsymbol{t} \cdot \hat{\boldsymbol{u}}\, d\Gamma
\end{equation}
and the minimum mean compliance problem under a volume constraint is formulated as follows:
\begin{equation}\label{eq:3-8}
\begin{aligned}
& \min_{\psi} \quad l(\hat{\boldsymbol{u}}) \\
& \text{s.t.} \quad
\begin{cases}
\bar z = H_{\beta,\eta}(\tilde\rho), \quad \tilde\rho = \mathcal{F}(\psi) & \text{\eqref{eq:3-1}, \eqref{eq:3-2}} \\
\text{static equilibrium \eqref{eq:3-6}} \\
\bar V := \dfrac{1}{|\Omega|}\displaystyle\int_\Omega \bar z\, dV \leq V_{\max} \\
0 \leq \psi \leq 1
\end{cases}
\end{aligned}
\end{equation}
where $\mathcal{F}$ is the filter operator of \eqref{eq:3-1} and $V_{\max}$ is the admissible volume fraction.

\subsection{Residual stress minimization problem}\label{sec:3-3}

Next, we formulate the optimization problem aimed at suppressing build failures caused by the residual stress during the building process. The quantity to be evaluated is the residual stress field $\boldsymbol{\sigma}^{(N)}$ at the completion of the building process obtained by the layer-by-layer analysis of Section~\ref{sec:2}.

In density-based topology optimization with a stress objective, the singularity problem is well known: the local stress remains finite in the limit of vanishing density, which produces discontinuities in the vicinity of the optimum \cite{DuysinxBendsoe1998}. To avoid this problem, we introduce a relaxed stress measure:
\begin{equation}\label{eq:3-9}
\hat\sigma(\boldsymbol{x}) = \frac{\eta_s(\bar z)}{\eta_E(\bar z)}\, \frac{\sigma_{vm}\bigl(\boldsymbol{\sigma}^{(N)}\bigr)}{\sigma_{\max}},
\qquad
\eta_s(\bar z) = \epsilon_E + (1 - \epsilon_E)\, \bar z^{\, q_s}
\end{equation}
Here, $\sigma_{vm}(\boldsymbol{\sigma}^{(N)})/\eta_E(\bar z)$ corresponds to the local stress amplitude from which the apparent stress reduction due to the interpolated stiffness has been removed, and it remains finite in the limit of vanishing density. Multiplying it by the relaxation coefficient $\eta_s(\bar z)$ with $0 < q_s < 1$, the relaxed stress measure $\hat\sigma$ smoothly degenerates to zero in the void limit, and the singularity is avoided by this qp relaxation \cite{Bruggi2008}. In addition, since $q_s < 1$ leads to $\eta_s/\eta_E > 1$ at intermediate densities, the stress is evaluated relatively larger there, which also contributes to the suppression of intermediate densities. The reference stress $\sigma_{\max}$ is used for normalization, and the initial yield stress $\sigma_{y1}$ of the solid material is used in this study.

To handle the suppression of the maximum residual stress in a differentiable form, the P-norm aggregation \cite{Le2010} of $\hat\sigma$ is used as the objective function:
\begin{equation}\label{eq:3-10}
J(\bar z) = \left( \int_\Omega \hat\sigma^{\, p}\, dV \right)^{1/p}
\end{equation}
where $p$ is the aggregation exponent, and $J$ approaches the maximum of $\hat\sigma$ in the limit $p \to \infty$. The residual stress minimization problem is formulated as follows:
\begin{equation}\label{eq:3-11}
\begin{aligned}
& \min_{\psi} \quad J(\bar z) \\
& \text{s.t.} \quad
\begin{cases}
\bar z = H_{\beta,\eta}(\tilde\rho), \quad \tilde\rho = \mathcal{F}(\psi) \\
\text{design-dependent layer-by-layer analysis \eqref{eq:2-18}} \quad n = 1, \ldots, N \\
\bar V \leq V_{\max} \\
0 \leq \psi \leq 1
\end{cases}
\end{aligned}
\end{equation}

\subsection{Residual stress minimization problem under a compliance constraint}\label{sec:3-4}

With the residual stress minimization problem \eqref{eq:3-11} alone, structures that are easy to build can be obtained at the cost of the stiffness required for the product. Indeed, since the inherent strain load acts only where material exists, without any constraint the optimization tends to remove material, that is, to degrade the final-use performance. In this study, we therefore consider the following optimization problem, which ensures the product performance by imposing an upper bound on the mean compliance while minimizing the residual stress to enhance the manufacturing reliability:
\begin{equation}\label{eq:3-12}
\begin{aligned}
& \min_{\psi} \quad J(\bar z) = \left( \int_\Omega \hat\sigma^{\, p}\, dV \right)^{1/p} \\
& \text{s.t.} \quad
\begin{cases}
\bar z = H_{\beta,\eta}(\tilde\rho), \quad \tilde\rho = \mathcal{F}(\psi) \\
\text{design-dependent layer-by-layer analysis \eqref{eq:2-18}} & n = 1, \ldots, N \\
\text{static equilibrium \eqref{eq:3-6}} \\
l(\hat{\boldsymbol{u}}) \leq C_{\max} \\
\bar V \leq V_{\max} \\
0 \leq \psi \leq 1
\end{cases}
\end{aligned}
\end{equation}
where $C_{\max}$ is the admissible upper bound of the mean compliance, which is set according to the required product stiffness. In the numerical examples of Section~\ref{sec:6}, it is given on the basis of the compliance of a reference design. A characteristic of problem \eqref{eq:3-12} is that it involves two kinds of state systems: the state equations of the building process, namely the layer-by-layer elastoplastic analysis with $N$ steps, and the state equation of the final use, namely static linear elasticity.

To solve problem \eqref{eq:3-12} by a gradient-based method, the design sensitivity of the objective function $J$ is required. Since $J$ depends on the design variable through the history-dependent elastoplastic analysis of $N$ steps, the efficient and exact evaluation of the sensitivity is an essential challenge. In the next section, the sensitivity analysis based on the adjoint method is formulated.

\section{Adjoint sensitivity analysis}\label{sec:4}

In this section, the design sensitivity of the objective function $J$ of the optimization problem \eqref{eq:3-12} is derived by the adjoint method. Since $J$ depends on the final state of the layer-by-layer elastoplastic analysis of $N$ steps, the sensitivity analysis requires a rigorous treatment of the history dependence over all the steps. The core of this section is that the adjoint system associated with the one-step recurrence structure \eqref{eq:2-14} of the state pair $(\boldsymbol{\sigma}^{(n)}, \bar\alpha^{(n)})$ presented in Section~\ref{sec:2} becomes a reverse recurrence that mirrors the forward analysis. That is, on the adjoint side, an ``adjoint pseudo-stress'' tensor field is carried over in the reverse order of the building sequence, and the coefficients of its transport and stiffness are all constructed by reusing the return-mapping projection $\mathbb{T}^{(n)}$ and the algorithmic tangent $\mathbb{D}^{ats\,(n)}$ of the forward analysis (Equations~\eqref{eq:2-15} and \eqref{eq:2-16}). Section~\ref{sec:4-1} prepares the derivatives of the update mappings and a basic identity, Section~\ref{sec:4-2} derives the adjoint equations from a Lagrangian, Section~\ref{sec:4-3} presents the reverse chain of the adjoint pseudo-stress as the implementation form, Section~\ref{sec:4-4} gives the final form of the design sensitivity, and Section~\ref{sec:4-5} discusses the degeneration in the elastic limit.

\subsection{Derivatives of the update mappings and a basic identity}\label{sec:4-1}

The coefficients of the adjoint equations are determined by the derivatives of the update mappings $A^{(n)}$ and $\boldsymbol{S}^{(n)}$ in \eqref{eq:2-14}. The derivative of the stress update with respect to the trial stress is the return-mapping projection $\mathbb{T}^{(n)}$ in \eqref{eq:2-15}, and the remaining derivatives are, in the plastic state ($\Delta\bar\alpha^{(n)} > 0$),
\begin{equation}\label{eq:4-1}
\frac{\partial \boldsymbol{S}^{(n)}}{\partial \bar\alpha^{(n-1)}} = \sqrt{\tfrac{2}{3}}\,\frac{3\mu H}{3\mu+H}\,\hat{\boldsymbol{n}}^{(n)},
\quad
\frac{\partial A^{(n)}}{\partial \boldsymbol{\sigma}^{tr}} = \frac{\sqrt{3/2}}{3\mu+H}\,\hat{\boldsymbol{n}}^{(n)},
\quad
\frac{\partial A^{(n)}}{\partial \bar\alpha^{(n-1)}} = \frac{3\mu}{3\mu+H}
\end{equation}
In the elastic state they are $\boldsymbol{0}$, $\boldsymbol{0}$, and $1$, respectively, with $\mathbb{T}^{(n)} = \mathbb{I}$. The elastic/plastic branch is determined by the active state of the KKT conditions \eqref{eq:2-10} of the forward analysis. To express this branch concisely in the following equations, we introduce the indicator function of the KKT active set,
\begin{equation*}
\chi_{\mathcal{P}}^{(n)}(\boldsymbol{x}) =
\begin{cases}
1 & \Delta\bar\alpha^{(n)}(\boldsymbol{x}) > 0 \\
0 & \text{otherwise}
\end{cases}
\end{equation*}
in the same manner as the layer indicator $\chi_{L_n}$ in \eqref{eq:2-4} and the activation indicator $\bar\chi^{(n)}$ in \eqref{eq:2-5}. Furthermore, the following basic identity holds between $\mathbb{T}^{(n)}$ and the algorithmic tangent $\mathbb{D}^{ats\,(n)}$:
\begin{equation}\label{eq:4-2}
\mathbb{T}^{(n)} = \mathbb{D}^{ats\,(n)} : (\mathbb{D}^{e})^{-1}, \qquad
\mathbb{D}^e : \mathbb{T}^{(n)T} = \mathbb{D}^{ats\,(n)}
\end{equation}
$\mathbb{T}^{(n)}$ is symmetric and commutes with $(\mathbb{D}^{e})^{-1}$ because they share the basis $\mathbb{I}^{vol}$, $\mathbb{I}^{dev}$, and $\hat{\boldsymbol{n}}\otimes\hat{\boldsymbol{n}}$. Equation~\eqref{eq:4-2} means that ``$\mathbb{T}^{(n)}$ is the algorithmic tangent measured in elastic units,'' and it is used repeatedly in the following derivation.

\subsection{Lagrangian and adjoint equations}\label{sec:4-2}

We treat the objective function as a functional of the physical density and the final stress, $J = J(\bar z, \boldsymbol{\sigma}^{(N)})$, and write its derivative with respect to the final stress as
\begin{equation}\label{eq:4-3}
\boldsymbol{f}_\sigma := \frac{\partial J}{\partial \boldsymbol{\sigma}^{(N)}}
\end{equation}
which is a deviatoric tensor field. For the objective function \eqref{eq:3-10} of Section~\ref{sec:3}, which is the P-norm of the relaxed stress $\hat\sigma = (\eta_s/\eta_E)\,\sigma_{vm}/\sigma_{\max}$, it reads
\begin{equation}\label{eq:4-4}
\boldsymbol{f}_\sigma
= \frac{3}{2}\, J^{1-p} \left( \frac{\eta_s(\bar z)}{\eta_E(\bar z)\, \sigma_{\max}} \right)^{p} \sigma_{vm}^{\,p-2}\bigl(\boldsymbol{\sigma}^{(N)}\bigr)\, \boldsymbol{s}^{(N)}
\end{equation}
As the adjoint fields, we introduce three multipliers: the global multiplier $\boldsymbol{\lambda}^{(n)} \in V$ associated with the equilibrium \eqref{eq:2-18}, the symmetric tensor field $\boldsymbol{\zeta}^{(n)}$ associated with the stress update, and the scalar field $\gamma^{(n)}$ associated with the hardening variable update. The Lagrangian is then defined as
\begin{equation}\label{eq:4-5}
\begin{aligned}
\mathcal{L} = J
&+ \sum_{n=1}^{N} \int_\Omega \bar\chi^{(n)}\, \boldsymbol{\sigma}^{(n)} : \boldsymbol{\varepsilon}\bigl(\boldsymbol{\lambda}^{(n)}\bigr)\, dV \\
&+ \sum_{n=1}^{N} \int_\Omega \boldsymbol{\zeta}^{(n)} : \bigl(\boldsymbol{\sigma}^{(n)} - \boldsymbol{S}^{(n)}\bigr)\, dV
+ \sum_{n=1}^{N} \int_\Omega \gamma^{(n)} \bigl(\bar\alpha^{(n)} - A^{(n)}\bigr)\, dV
\end{aligned}
\end{equation}
In \eqref{eq:4-5}, the state fields $(\boldsymbol{u}^{(n)}, \boldsymbol{\sigma}^{(n)}, \bar\alpha^{(n)})$ and the adjoint fields $(\boldsymbol{\lambda}^{(n)}, \boldsymbol{\zeta}^{(n)}, \gamma^{(n)})$ are all treated as mutually independent fields, following the so-called C\'ea's method \cite{Cea1986}. Setting the derivatives in the directions of the adjoint fields to zero recovers the constraints of the forward analysis, namely the equilibrium \eqref{eq:2-18} and the state update \eqref{eq:2-14}; at the solution of the forward analysis, all the constraints are satisfied and $\mathcal{L} = J$. Setting the derivatives in the directions of the state fields to zero, on the other hand, yields the adjoint equations derived below. What is essential here is that the dependency structure of this formulation is a genuine one-step recurrence: $\boldsymbol{S}^{(n+1)}$ and $A^{(n+1)}$ depend only on $(\boldsymbol{u}^{(n+1)}, \boldsymbol{\sigma}^{(n)}, \bar\alpha^{(n)})$, and the equilibrium \eqref{eq:2-18} depends only on $\boldsymbol{\sigma}^{(n)}$.

This one-step structure is a consequence of the choice of the state variables and does not hold for the standard choice. If the state is taken as the pair of the plastic strain and the hardening variable $(\boldsymbol{\varepsilon}^p, \bar\alpha)$, which is the standard choice for static elastoplastic problems \cite{Michaleris1994,Kato2015,Alberdi2018,Granlund2024}, and applied directly to the accumulated form \eqref{eq:2-8}, the carried-over stress $\boldsymbol{\sigma}^{(n-1)} = \mathbb{D}^e : [\sum_{k \leq n-1}(\boldsymbol{\varepsilon}(\boldsymbol{u}^{(k)}) - \chi_{L_k}\boldsymbol{\varepsilon}^*_0) - \boldsymbol{\varepsilon}^{p\,(n-1)}]$ depends on the displacements of all the steps; the dependency is then no longer a one-step recurrence, and the adjoint equations become coupled over all the steps. By including the accumulated stress itself in the state as in \eqref{eq:2-14}, the dependency becomes genuinely one-step, and the reverse recurrences below close.

To derive the adjoint equations, we write down the variation of $\mathcal{L}$ with respect to the state fields. First, the variations of the trial stress \eqref{eq:2-11} and the update mappings \eqref{eq:2-14} are expressed using $\mathbb{T}^{(n)}$ (Equation~\eqref{eq:2-15}) and the derivatives \eqref{eq:4-1} as
\begin{equation}\label{eq:4-6}
\begin{aligned}
&\delta\boldsymbol{\sigma}^{tr\,(n)} = \mathbb{D}^e : \boldsymbol{\varepsilon}\bigl(\delta\boldsymbol{u}^{(n)}\bigr) + \delta\boldsymbol{\sigma}^{(n-1)}, \quad
\delta\boldsymbol{S}^{(n)} = \mathbb{T}^{(n)} : \delta\boldsymbol{\sigma}^{tr\,(n)} + \frac{\partial \boldsymbol{S}^{(n)}}{\partial \bar\alpha^{(n-1)}}\,\delta\bar\alpha^{(n-1)}, \\
&\delta A^{(n)} = \frac{\partial A^{(n)}}{\partial \boldsymbol{\sigma}^{tr}} : \delta\boldsymbol{\sigma}^{tr\,(n)} + \frac{\partial A^{(n)}}{\partial \bar\alpha^{(n-1)}}\,\delta\bar\alpha^{(n-1)}
\end{aligned}
\end{equation}
In particular, since the trial stress depends identically on the stress of the previous step ($\partial\boldsymbol{\sigma}^{tr\,(n)}/\partial\boldsymbol{\sigma}^{(n-1)} = \mathbb{I}$), we have $\partial\boldsymbol{S}^{(n+1)}/\partial\boldsymbol{\sigma}^{(n)} = \mathbb{T}^{(n+1)}$.

Substituting \eqref{eq:4-6} into the variation of \eqref{eq:4-5} and collecting terms for each variation $\delta\boldsymbol{\sigma}^{(n)}$ and $\delta\bar\alpha^{(n)}$, we relabel the summation index by shifting the sums containing $n+1$ to $n$, in order to pick up $\delta\boldsymbol{\sigma}^{(n)}$ and $\delta\bar\alpha^{(n)}$ appearing in $\delta\boldsymbol{S}^{(n+1)}$ and $\delta A^{(n+1)}$. This operation is the discrete counterpart of the integration by parts in time in the adjoint derivation of time-continuous systems. After rearrangement, the variation of $\mathcal{L}$ is organized into the following three groups:
\begin{equation}\label{eq:4-7}
\begin{aligned}
\delta\mathcal{L}
= & \sum_{n=1}^{N} \int_\Omega \left[ \delta_{nN}\,\boldsymbol{f}_\sigma + \bar\chi^{(n)}\boldsymbol{\varepsilon}\bigl(\boldsymbol{\lambda}^{(n)}\bigr) + \boldsymbol{\zeta}^{(n)}
- \mathbb{T}^{(n+1)T} : \boldsymbol{\zeta}^{(n+1)} \right. \\
& \hspace{9.5em} \left. - \gamma^{(n+1)}\,\frac{\partial A^{(n+1)}}{\partial \boldsymbol{\sigma}^{tr}} \right] : \delta\boldsymbol{\sigma}^{(n)}\, dV \\
& + \sum_{n=1}^{N} \int_\Omega \left[ \gamma^{(n)} - \boldsymbol{\zeta}^{(n+1)} : \frac{\partial \boldsymbol{S}^{(n+1)}}{\partial \bar\alpha^{(n)}} - \gamma^{(n+1)}\,\frac{\partial A^{(n+1)}}{\partial \bar\alpha^{(n)}} \right] \delta\bar\alpha^{(n)}\, dV \\
& - \sum_{n=1}^{N} \int_\Omega \left[ \mathbb{T}^{(n)T} : \boldsymbol{\zeta}^{(n)} + \gamma^{(n)}\,\frac{\partial A^{(n)}}{\partial \boldsymbol{\sigma}^{tr}} \right] : \mathbb{D}^e : \boldsymbol{\varepsilon}\bigl(\delta\boldsymbol{u}^{(n)}\bigr)\, dV
\end{aligned}
\end{equation}
where we set $\boldsymbol{\zeta}^{(N+1)} := \boldsymbol{0}$ and $\gamma^{(N+1)} := 0$ so that the sums are unified to $n = 1, \ldots, N$. In the first group, $\delta_{nN}\boldsymbol{f}_\sigma$ is the variation of the objective function, which is present only for $n = N$, $\bar\chi^{(n)}\boldsymbol{\varepsilon}(\boldsymbol{\lambda}^{(n)})$ is the variation of the equilibrium equation \eqref{eq:2-18}, $\boldsymbol{\zeta}^{(n)}$ is that of the stress-update equation, and the remaining two terms are the coupling terms transferred from the update mappings of the next step by the index relabeling.

The stationarity condition requires $\delta\mathcal{L} = 0$ for arbitrary variations of the symmetric tensor field $\delta\boldsymbol{\sigma}^{(n)}$, the scalar field $\delta\bar\alpha^{(n)}$, and $\delta\boldsymbol{u}^{(n)} \in V$. Equivalently, each bracket in \eqref{eq:4-7} must vanish at each step and at each material point. Setting the coefficient of the first group to zero, solving for $\boldsymbol{\zeta}^{(n)}$, and substituting the explicit derivatives \eqref{eq:4-1}, we obtain the reverse recurrence for $\boldsymbol{\zeta}$:
\begin{equation}\label{eq:4-8}
\boldsymbol{\zeta}^{(n)} = -\,\delta_{nN}\, \boldsymbol{f}_\sigma
- \bar\chi^{(n)} \boldsymbol{\varepsilon}\bigl(\boldsymbol{\lambda}^{(n)}\bigr)
+ \mathbb{T}^{(n+1)T} : \boldsymbol{\zeta}^{(n+1)}
+ \chi_{\mathcal{P}}^{(n+1)}\, \gamma^{(n+1)}\, \frac{\sqrt{3/2}}{3\mu+H}\, \hat{\boldsymbol{n}}^{(n+1)}
\end{equation}
Equation~\eqref{eq:4-8} is evaluated in the reverse order $n = N, N-1, \ldots, 1$. Since the plastic gate $\chi_{\mathcal{P}}^{(n+1)}$ has already been fixed by the active set of the KKT conditions \eqref{eq:2-10} in the forward analysis, the adjoint system is a linear problem along the path of the forward analysis, and the reverse sweep requires no nonlinear iteration. Similarly, setting the coefficient of the second group of \eqref{eq:4-7} to zero yields the reverse recurrence for $\gamma$:
\begin{equation}\label{eq:4-9}
\gamma^{(n)} = \chi_{\mathcal{P}}^{(n+1)}\, \sqrt{\tfrac{2}{3}}\,\frac{3\mu H}{3\mu+H}\,\bigl(\hat{\boldsymbol{n}}^{(n+1)}:\boldsymbol{\zeta}^{(n+1)}\bigr)
+ \Bigl( 1 - \chi_{\mathcal{P}}^{(n+1)}\, \frac{H}{3\mu+H} \Bigr)\, \gamma^{(n+1)}
\end{equation}
That is, if step $n+1$ is plastic, the coefficient of the second term becomes $3\mu/(3\mu+H)$, and if it is elastic, the recurrence degenerates to the simple carry-over $\gamma^{(n)} = \gamma^{(n+1)}$.

The terminal condition follows naturally: the $n = N$ component of the second group of \eqref{eq:4-7} gives $\gamma^{(N)} = 0$ owing to $\boldsymbol{\zeta}^{(N+1)} = \boldsymbol{0}$ and $\gamma^{(N+1)} = 0$. This is a consequence of the fact that $J$ does not depend explicitly on $\bar\alpha$. Finally, the vanishing of the coefficient of the third group of \eqref{eq:4-7} is equivalent to a weak form for arbitrary $\delta\boldsymbol{u}^{(n)} \in V$. Rewriting the integrand as $\mathbb{D}^e : \bigl[\mathbb{T}^{(n)T} : \boldsymbol{\zeta}^{(n)} + \gamma^{(n)}\,\partial A^{(n)}/\partial\boldsymbol{\sigma}^{tr}\bigr] : \boldsymbol{\varepsilon}(\delta\boldsymbol{u}^{(n)})$ and using the identity $\mathbb{D}^e : \mathbb{T}^{(n)T} = \mathbb{D}^{ats\,(n)}$ in \eqref{eq:4-2} and $\mathbb{D}^e : \hat{\boldsymbol{n}} = 2\mu\hat{\boldsymbol{n}}$, which holds because $\hat{\boldsymbol{n}}$ is deviatoric, we obtain the weak-form equation for $\boldsymbol{\lambda}^{(n)}$:
\begin{equation}\label{eq:4-10}
\int_\Omega \Bigl[ \mathbb{D}^{ats\,(n)} : \boldsymbol{\zeta}^{(n)}
+ \chi_{\mathcal{P}}^{(n)}\,\gamma^{(n)}\, \frac{\sqrt{3/2}\cdot 2\mu}{3\mu+H}\, \hat{\boldsymbol{n}}^{(n)} \Bigr] : \boldsymbol{\varepsilon}(\boldsymbol{v})\, dV = 0
\qquad \forall\, \boldsymbol{v} \in V
\end{equation}

\subsection{Reverse carry-over of the adjoint pseudo-stress}\label{sec:4-3}

The adjoint system \eqref{eq:4-8}--\eqref{eq:4-10} can be recast into an implementation form having the same structure as the forward analysis, that is, an initial stress plus a state carry-over, by introducing the \emph{adjoint pseudo-stress}, which converts $\boldsymbol{\zeta}$ into a ``stress in elastic units'':
\begin{equation}\label{eq:4-11}
\boldsymbol{P}^{(n)} := -\,\mathbb{D}^e : \boldsymbol{\zeta}^{(n)}
\end{equation}
Applying $-\mathbb{D}^e$ to \eqref{eq:4-8} and using the commutativity in \eqref{eq:4-2} and $\mathbb{D}^e : \hat{\boldsymbol{n}} = 2\mu\hat{\boldsymbol{n}}$, we obtain
\begin{equation}\label{eq:4-12}
\begin{aligned}
\boldsymbol{P}^{(n)} &= \bar\chi^{(n)}\, \mathbb{D}^e : \boldsymbol{\varepsilon}\bigl(\boldsymbol{\lambda}^{(n)}\bigr) + \boldsymbol{Q}^{(n)}, \\
\boldsymbol{Q}^{(n)} &:= \mathbb{T}^{(n+1)} : \boldsymbol{P}^{(n+1)}
- \chi_{\mathcal{P}}^{(n+1)}\, \gamma^{(n+1)}\, \frac{\sqrt{3/2}\cdot 2\mu}{3\mu+H}\, \hat{\boldsymbol{n}}^{(n+1)}
+ \delta_{nN}\, \mathbb{D}^e : \boldsymbol{f}_\sigma
\end{aligned}
\end{equation}
Substituting $\boldsymbol{\zeta}^{(n)} = -(\mathbb{D}^{e})^{-1} : \boldsymbol{P}^{(n)}$ into the $\boldsymbol{\lambda}$ equation \eqref{eq:4-10} yields
\begin{equation}\label{eq:4-13}
\int_\Omega \bar\chi^{(n)}\, \boldsymbol{\varepsilon}(\boldsymbol{v}) : \mathbb{D}^{ats\,(n)} : \boldsymbol{\varepsilon}\bigl(\boldsymbol{\lambda}^{(n)}\bigr)\, dV
= -\int_\Omega \boldsymbol{\Sigma}^{il\,(n)} : \boldsymbol{\varepsilon}(\boldsymbol{v})\, dV
\qquad \forall\, \boldsymbol{v} \in V
\end{equation}
\begin{equation}\label{eq:4-14}
\boldsymbol{\Sigma}^{il\,(n)} = \mathbb{T}^{(n)} : \boldsymbol{Q}^{(n)}
- \chi_{\mathcal{P}}^{(n)}\, \gamma^{(n)}\, \frac{\sqrt{3/2}\cdot 2\mu}{3\mu+H}\, \hat{\boldsymbol{n}}^{(n)}
\end{equation}
That is, the $\boldsymbol{\lambda}$ equation of each reverse step is a linear elasticity-type problem with the algorithmic tangent $\mathbb{D}^{ats\,(n)}$ of forward step $n$ as the stiffness and $\boldsymbol{\Sigma}^{il\,(n)}$ as the initial stress, and it has the same structure as the forward equilibrium \eqref{eq:2-18}. This correspondence is summarized in Table~\ref{tab:4-1}.

\begin{table}[pos=!htbp]
  \centering
  \small
  \caption{Mirror symmetry between the forward analysis and the adjoint analysis}
  \label{tab:4-1}
  \begin{tabularx}{\textwidth}{>{\raggedright\arraybackslash}p{3.1cm} >{\raggedright\arraybackslash}X >{\raggedright\arraybackslash}X}
    \toprule
    & Forward analysis ($n = 1 \to N$) & Adjoint analysis ($n = N \to 1$) \\
    \midrule
    Global field & $\boldsymbol{u}^{(n)}$ & $\boldsymbol{\lambda}^{(n)}$ \\
    Carried-over stress state & $\boldsymbol{\sigma}^{(n-1)}$ & $\boldsymbol{P}^{(n+1)}$ \\
    Carried-over local scalar state & $\bar\alpha^{(n-1)}$ & $\gamma^{(n+1)}$ \\
    Driving term & inherent strain $\chi_{L_n}\boldsymbol{\varepsilon}^*_0$ (every step) & objective load $\mathbb{D}^e : \boldsymbol{f}_\sigma$ (final step only) \\
    Stiffness within a step & $\mathbb{D}^{ats\,(n)}$ & $\mathbb{D}^{ats\,(n)}$ (identical; no transpose is needed owing to the major symmetry of $\mathbb{D}^{ats}$) \\
    Transport between steps & identity addition to the trial stress \eqref{eq:2-11} & return-mapping projection $\mathbb{T}^{(n)}$ \eqref{eq:4-14} \\
    \bottomrule
  \end{tabularx}
\end{table}

The procedure of the reverse sweep is presented in Algorithm~\ref{alg:2}.

\begin{algorithm}[tb]
\caption{Reverse sweep of the adjoint pseudo-stress}
\label{alg:2}
\begin{algorithmic}[1]
\State $n = N$: $\boldsymbol{Q}^{(N)} = \mathbb{D}^e : \boldsymbol{f}_\sigma$, $\gamma^{(N)} = 0$ \Comment{objective load injected only at the final step}
\For{$n = N, N-1, \ldots, 1$}
  \State Construct the adjoint initial stress $\boldsymbol{\Sigma}^{il\,(n)}$ by \eqref{eq:4-14} and solve the $\boldsymbol{\lambda}$ equation \eqref{eq:4-13} to obtain $\boldsymbol{\lambda}^{(n)}$, where both the stiffness $\mathbb{D}^{ats\,(n)}$ and the projection $\mathbb{T}^{(n)}$ are constructed from the stored quantities of forward step $n$
  \State Evaluate $\boldsymbol{P}^{(n)}$ by \eqref{eq:4-12} and carry
  \Statex \hspace{\algorithmicindent}\hspace{\algorithmicindent} $\gamma^{(n-1)} = -\chi_{\mathcal{P}}^{(n)}\sqrt{\tfrac{3}{2}}\,\dfrac{H}{3\mu+H}\bigl(\hat{\boldsymbol{n}}^{(n)}:\boldsymbol{P}^{(n)}\bigr) + \Bigl( 1 - \chi_{\mathcal{P}}^{(n)}\dfrac{H}{3\mu+H} \Bigr)\gamma^{(n)}$
  \Statex \hspace{\algorithmicindent} over to the next reverse step
\EndFor
\end{algorithmic}
\end{algorithm}

The essential points of this formulation are as follows. First, the objective load is injected only at the final step, and thereafter the chain of $\boldsymbol{P}$ automatically accounts for the transport by the $\mathbb{T}$ projections. Second, the coefficient tensors $\mathbb{D}^{ats\,(n)}$, $\mathbb{T}^{(n)}$, $\hat{\boldsymbol{n}}^{(n)}$, $\theta^{(n)}$, and $\bar\theta^{(n)}$ appearing in the adjoint analysis are all quantities that have already been constructed and stored in the return mapping of the forward analysis, Equations~\eqref{eq:2-15} and \eqref{eq:2-16}; no new coefficients need to be constructed on the adjoint side. Consequently, the adjoint fields can be constructed with the same layer-by-layer structure as the physical fields, namely an initial stress plus a carry-over of state variables, and the implementation becomes analogous to the forward analysis. This is an extension, to the accumulated form of AM, of the adjoint structure established for static elastoplastic problems, in which the reverse scheme mirrors the forward state update \cite{Alberdi2018,Granlund2024}.

\subsection{Final form of the design sensitivity}\label{sec:4-4}

\subsubsection{Parameterization of the sensitivity}\label{sec:4-4-1}

Once the adjoint fields are determined, the design sensitivity is given by collecting the explicit derivatives of $\mathcal{L}$ with respect to the physical density $\bar z$, that is, the derivatives taken with the state fields held fixed. Care must be taken here because the meaning of ``holding the state fields fixed'' depends on what the implementation of the forward analysis carries over as the state. The multiplier fields $(\boldsymbol{\lambda}, \boldsymbol{P}, \gamma)$ themselves are unique quantities independent of the parameterization, but the split into explicit $\bar z$-derivatives is determined by what is held fixed when $\bar z$ is perturbed. The implementation in this study carries over the total strain, the plastic strain, and $\bar\alpha$ as the state between steps, and reconstructs the inherited stress at each step as $\boldsymbol{\sigma}^{(n-1)} = \mathbb{D}^e(\bar z) : \boldsymbol{\varepsilon}^{e\,(n-1)}$ with the current $\bar z$. We refer to this as the strain-state parameterization. A perturbation of $\bar z$ therefore propagates into the inherited stress through $\mathbb{D}^e(\bar z)$. This differs, in the explicit $\bar z$-derivatives, from a parameterization that carries over the stress values as fixed; the solutions are identical, but the sensitivity expressions differ. The difference between the two degenerates and is not observable in elastic problems, as shown in Section~\ref{sec:4-5}, and becomes manifest only in elastoplasticity. The sensitivity expressions derived below correspond to the strain-state parameterization, and the finite difference verification in Section~\ref{sec:5} is also performed under this correspondence.

In the strain-state parameterization, the governing equations are expressed in the strain variables, and the multipliers correspond as follows: $\boldsymbol{\lambda}^{(n)}$ remains the multiplier of the equilibrium equation \eqref{eq:2-18} and $\gamma^{(n)}$ that of the update equation of $\bar\alpha$, and the multiplier of the update equation of the plastic strain, $\Delta\boldsymbol{\varepsilon}^{p\,(n)} = \sqrt{3/2}\,\Delta\bar\alpha^{(n)}\hat{\boldsymbol{n}}^{(n)}$, is given by $\boldsymbol{P}^{(n)}$. The latter can be seen as follows: injecting an offset $\delta\boldsymbol{\varepsilon}^p$ into the plastic strain $\boldsymbol{\varepsilon}^{p\,(n)}$ in the strain-variable system changes the stress by $-\mathbb{D}^e : \delta\boldsymbol{\varepsilon}^p$ with the strains held fixed; this is equivalent to an offset $-\mathbb{D}^e : \delta\boldsymbol{\varepsilon}^p$ to the stress update mapping $\boldsymbol{S}^{(n)}$, and the corresponding change of $J$ is $dJ = -\boldsymbol{\zeta}^{(n)}:(-\mathbb{D}^e : \delta\boldsymbol{\varepsilon}^p) = -\boldsymbol{P}^{(n)}:\delta\boldsymbol{\varepsilon}^p$. Hence the multiplier of the update equation of $\boldsymbol{\varepsilon}^p$ is $\boldsymbol{P}^{(n)}$.

\subsubsection{Explicit derivatives and the final form}\label{sec:4-4-2}

We organize the explicit $\bar z$-derivatives of each quantity in the strain-state parameterization. Since the stress is proportional to $\eta_E$ in the form $\boldsymbol{\sigma}^{(n)} = \mathbb{D}^e(\bar z) : \boldsymbol{\varepsilon}^{e\,(n)}$,
\begin{equation}\label{eq:4-15}
\frac{\partial \boldsymbol{\sigma}^{(n)}}{\partial \bar z}\bigg|_{\varepsilon}
= \frac{\eta_E'(\bar z)}{\eta_E(\bar z)}\, \boldsymbol{\sigma}^{(n)}
\end{equation}
where $\eta_E'$ is the derivative of the RAMP interpolation \eqref{eq:3-3}. For the plastic increment, using $\partial\sigma_{vm}^{tr}/\partial\bar z|_\varepsilon = (\eta_E'/\eta_E)\sigma_{vm}^{tr}$, the update formula \eqref{eq:2-12}, and the consistency relation $\sigma_{vm}^{tr} = \sigma_y(\bar\alpha^{(n)}) + 3\mu\Delta\bar\alpha^{(n)}$, the terms proportional to $3\mu\Delta\bar\alpha$ cancel, and we obtain
\begin{equation}\label{eq:4-16}
\frac{\partial \Delta\bar\alpha^{(n)}}{\partial \bar z}\bigg|_{\varepsilon}
= \frac{1}{3\mu+H}\left[ \sigma_y\bigl(\bar\alpha^{(n)}\bigr)\, \frac{\eta_E'}{\eta_E}
- \eta_p'\, \sigma_{y1} - \eta_p'\, H_1\, \bar\alpha^{(n)} \right]
\end{equation}
where $\eta_p'$ is the derivative of the power-law interpolation \eqref{eq:3-4}. At elastic points the whole expression vanishes because $\Delta\bar\alpha^{(n)} = 0$. Note that the flow direction $\hat{\boldsymbol{n}}^{(n)}$ equals the direction of the deviatoric trial strain and does not depend explicitly on $\bar z$ because the elastic moduli cancel in the normalization. Hence $\partial(\Delta\boldsymbol{\varepsilon}^{p\,(n)})/\partial\bar z|_\varepsilon = \sqrt{3/2}\,(\partial\Delta\bar\alpha^{(n)}/\partial\bar z)\,\hat{\boldsymbol{n}}^{(n)}$.

The explicit derivative of the objective function takes a concise form as the logarithmic derivative of the relaxed stress \eqref{eq:3-9}. In $\hat\sigma = (\eta_s/\eta_E)\sigma_{vm}/\sigma_{\max}$, the contribution $\eta_E'/\eta_E$ of $\sigma_{vm}$ from \eqref{eq:4-15} and the contribution $\eta_s'/\eta_s - \eta_E'/\eta_E$ of the relaxation coefficient combine so that the $\eta_E$ dependence cancels:
\begin{equation*}
\frac{\partial \hat\sigma}{\partial \bar z}\bigg|_{\varepsilon} = \frac{\eta_s'(\bar z)}{\eta_s(\bar z)}\, \hat\sigma
\end{equation*}
Collecting the above, the final form of the design sensitivity, expressed as a sensitivity density field, is given by
\begin{equation}\label{eq:4-17}
\begin{aligned}
\frac{dJ}{d\bar z} =\ & \underbrace{J^{1-p}\, \hat\sigma^{\,p}\, \frac{\eta_s'}{\eta_s}}_{\text{\circled{1} explicit term}}
\ +\ \underbrace{\sum_{n=1}^{N} \bar\chi^{(n)}\, \frac{\eta_E'}{\eta_E}\, \boldsymbol{\sigma}^{(n)} : \boldsymbol{\varepsilon}\bigl(\boldsymbol{\lambda}^{(n)}\bigr)}_{\text{\circled{2} adjoint term}} \\
& -\ \underbrace{\sum_{n=1}^{N} \chi_{\mathcal{P}}^{(n)}
\left( \sqrt{\tfrac{3}{2}}\,\bigl(\hat{\boldsymbol{n}}^{(n)}:\boldsymbol{P}^{(n)}\bigr) + \gamma^{(n)} \right)
\frac{\partial \Delta\bar\alpha^{(n)}}{\partial \bar z}\bigg|_{\varepsilon}}_{\text{\circled{3} plastic term}}
\end{aligned}
\end{equation}
The properties of each term are as follows. Term \circled{1} is the explicit derivative of the relaxed objective function \eqref{eq:3-10} and is localized in the stress concentration regions of the final state. Term \circled{2} originates from the explicit derivative of the equilibrium equation \eqref{eq:2-18} and takes the form of the pair of the accumulated stress $\boldsymbol{\sigma}^{(n)}$ and the adjoint strain $\boldsymbol{\varepsilon}(\boldsymbol{\lambda}^{(n)})$ of the same step. In the inactive domain, this term automatically vanishes because $\boldsymbol{\sigma}^{(n)}$ is proportional to the very small stiffness of the quiet elements. Term \circled{3} originates from the explicit derivatives of the update equation of the plastic strain, whose multiplier is $\boldsymbol{P}^{(n)}$ as discussed in Section~\ref{sec:4-4-1}, and of the update equation of $\bar\alpha$, whose multiplier is $\gamma^{(n)}$, and its gate coincides with the active set of the KKT conditions \eqref{eq:2-10}.

\subsubsection{Chain through the filter and projection, and constraint sensitivities}\label{sec:4-4-3}

The sensitivity with respect to the design variable $\psi$ is given by the chain through the projection \eqref{eq:3-2} and the filter \eqref{eq:3-1}:
\begin{equation}\label{eq:4-18}
\frac{dJ}{d\psi} = \mathcal{F}^*\!\left[ H_{\beta,\eta}'(\tilde\rho)\, \frac{dJ}{d\bar z} \right], \qquad
H_{\beta,\eta}'(\tilde\rho) = \frac{\beta\,\bigl(1 - \tanh^2(\beta(\tilde\rho - \eta))\bigr)}{\tanh(\beta\eta) + \tanh(\beta(1-\eta))}
\end{equation}
where $\mathcal{F}^*$ is the adjoint of the filter operator $\mathcal{F}$; for the normalized cone kernel \eqref{eq:3-1}, it acts on the sensitivity field with the same kernel normalized on the evaluation-point side; the discrete form is given in Section~\ref{sec:5}.

The sensitivities of the constraint functions are standard. The volume constraint has the explicit form $d\bar V/d\bar z = 1/|\Omega|$, and the compliance constraint, owing to self-adjointness, is given by
\begin{equation}\label{eq:4-19}
\frac{dl}{d\bar z} = -\,\frac{\eta_E'(\bar z)}{\eta_E(\bar z)}\, \mathbb{D}^e(\bar z) : \boldsymbol{\varepsilon}(\hat{\boldsymbol{u}}) : \boldsymbol{\varepsilon}(\hat{\boldsymbol{u}})
\end{equation}
where $\hat{\boldsymbol{u}}$ is the solution of the static state \eqref{eq:3-6}; the derivation is standard and is omitted. The chain \eqref{eq:4-18} is applied to these sensitivities as well.

\subsection{Degeneration in the elastic limit}\label{sec:4-5}

When all points remain elastic, $\mathbb{T}^{(n)} = \mathbb{I}$ and $\gamma \equiv 0$, and \eqref{eq:4-12} degenerates to
\begin{equation}\label{eq:4-20}
\boldsymbol{P}^{(n)} = \mathbb{D}^e : \bigl(\boldsymbol{\eta}^{(n)} + \boldsymbol{f}_\sigma\bigr), \qquad
\boldsymbol{\eta}^{(n)} := \sum_{k=n}^{N} \bar\chi^{(k)}\boldsymbol{\varepsilon}\bigl(\boldsymbol{\lambda}^{(k)}\bigr)
\end{equation}
and the $\boldsymbol{\lambda}$ equation \eqref{eq:4-13} reduces to the form using the accumulated adjoint strain $\boldsymbol{\eta}$:
\begin{equation}\label{eq:4-21}
\int_\Omega \bar\chi^{(n)}\, \mathbb{D}^e : \bigl[\boldsymbol{\varepsilon}(\boldsymbol{\lambda}^{(n)}) + \boldsymbol{\eta}^{(n+1)} + \boldsymbol{f}_\sigma\bigr] : \boldsymbol{\varepsilon}(\boldsymbol{v})\, dV = 0
\qquad \forall\, \boldsymbol{v} \in V
\end{equation}
This coincides exactly with the adjoint equation of the linear elastic AM model. In this limit, the objective load $\boldsymbol{f}_\sigma$ is transported inside $\boldsymbol{P}$ without deformation ($\mathbb{T} = \mathbb{I}$), and hence it appears as a load at every step in \eqref{eq:4-21}. When plasticity occurs, $\boldsymbol{f}_\sigma$ is transported while being deformed by the intermediate projections $\mathbb{T}^{(k)}$ ($n < k \leq N$); such a simplification then no longer holds, and the transport by the $\boldsymbol{P}$ chain becomes essential.

As for the sensitivity, in the elastic limit \circled{3} $= 0$, and by the Abel summation-by-parts identity
\begin{equation}\label{eq:4-22}
\sum_{n=1}^{N} \boldsymbol{\sigma}^{(n)} : \boldsymbol{\varepsilon}\bigl(\boldsymbol{\lambda}^{(n)}\bigr)
= \sum_{n=1}^{N} \Delta\boldsymbol{\sigma}^{(n)} : \boldsymbol{\eta}^{(n)}, \qquad
\Delta\boldsymbol{\sigma}^{(n)} = \boldsymbol{\sigma}^{(n)} - \boldsymbol{\sigma}^{(n-1)}
\end{equation}
the sum \circled{1} $+$ \circled{2} coincides exactly with the sensitivity of the linear elastic AM model.
In the next section, the numerical implementation of the adjoint sensitivity of this section and its verification by finite differences are presented.

\section{Numerical implementation and sensitivity verification}\label{sec:5}

In this section, we present the numerical implementation of the adjoint sensitivity derived in Section~\ref{sec:4} and its verification by finite differences. Although the formulation up to Section~\ref{sec:4} has been written in the continuous weak-form setting, the actual sensitivities are evaluated for the discretized forward analysis. The condition for the two to be consistent, that is, at which evaluation points and with which quadrature rule the adjoint fields and the sensitivity integrands are computed, is of practical importance in determining the accuracy of the sensitivities. Section~\ref{sec:5-1} describes the finite element discretization, the discrete forms of the filter and projection, the implementation structure of the adjoint analysis, and the overall optimization algorithm. Section~\ref{sec:5-2} verifies the sensitivity expression \eqref{eq:4-17} by direct comparison with central finite differences. Section~\ref{sec:5-3} evaluates the computational complexity and discusses the advantage of the proposed method over existing methods.

\subsection{Numerical implementation}\label{sec:5-1}

\subsubsection{Discretization and discrete forms of the filter and projection}\label{sec:5-1-1}

The fixed design domain $\Omega$ is divided by a structured grid and discretized with quadrilateral elements using quadratic interpolation for the displacement field. The design variable $\psi$ takes an element-wise constant value $\psi_e$ ($e = 1, 2, \ldots, N_{el}$), and the material interpolation is evaluated at the element centers. The stress and the plastic state variables are stored at the Gauss points of each element.

The discrete form of the density filter \eqref{eq:3-1} is the weighted average with the element centers $\boldsymbol{x}_e$ as the evaluation points:
\begin{equation}\label{eq:5-1}
\tilde\rho_e = \frac{\displaystyle\sum_{i \in \mathcal{N}_e} w_{ei}\, \psi_i}{\displaystyle\sum_{i \in \mathcal{N}_e} w_{ei}},
\quad
w_{ei} = \max\bigl(0,\; r - \|\boldsymbol{x}_e - \boldsymbol{x}_i\|\bigr),
\quad
\mathcal{N}_e = \bigl\{\, i \;:\; \|\boldsymbol{x}_e - \boldsymbol{x}_i\| \leq r \,\bigr\}
\end{equation}
The weights $w_{ei} = w_{ie}$ are symmetric, and the filter operator is represented as a symmetric sparse matrix normalized by the row sums. The physical density is $\bar z_e = H_{\beta,\eta}(\tilde\rho_e)$ with the projection \eqref{eq:3-2}.

Owing to this symmetry, the discrete form of the sensitivity chain \eqref{eq:4-18} becomes
\begin{equation}\label{eq:5-2}
\frac{dJ}{d\psi_i}
= \sum_{e \in \mathcal{N}_i} \frac{w_{ei}}{\displaystyle\sum_{j \in \mathcal{N}_e} w_{ej}}\,
H_{\beta,\eta}'(\tilde\rho_e)\, \frac{dJ}{d\bar z_e}
\end{equation}
That is, the filter adjoint $\mathcal{F}^*$ is the operation of applying the transposed forward weight matrix with the normalization kept on the evaluation-point side $e$, not the transpose of the whole matrix including the normalization. The same chain is applied to the sensitivity of the compliance constraint \eqref{eq:4-19}.

A remark on discrete consistency is in order here. The adjoint equation \eqref{eq:4-13} and the sensitivity expression \eqref{eq:4-17} of Section~\ref{sec:4} are derived in the continuous setting, but they are meaningful only when the adjoint fields and the sensitivity integrands are computed at the same evaluation points and with the same quadrature rule as the forward analysis. The quantities $\boldsymbol{\sigma}^{(n)}$, $\hat{\boldsymbol{n}}^{(n)}$, and $\Delta\bar\alpha^{(n)}$ appearing in the sensitivity expression are all quantities stored by the forward analysis at the Gauss points; if they are evaluated at other points through interpolation or reconstruction, the resulting discrepancy directly appears as a sensitivity error. In the present implementation, the element integrations of the forward state variables, the adjoint variables, and the sensitivities are all aligned to the $3 \times 3$ Gauss points per element. If this alignment is broken, the agreement with finite differences remains at the level of the quadrature error ($10^{-2}$--$10^{-3}$) and does not improve with smaller step sizes. The agreement achieved when the alignment is maintained is shown in Section~\ref{sec:5-2}.

\subsubsection{Implementation structure of the adjoint analysis}\label{sec:5-1-2}

Each step of the reverse sweep of Algorithm~\ref{alg:2} in Section~\ref{sec:4} consists of the following two stages. First, the adjoint pseudo-stress $\boldsymbol{P}^{(n)}$ and the hardening multiplier $\gamma^{(n-1)}$ are updated by the recurrence \eqref{eq:4-12}; this is an algebraic operation closed at each Gauss point and requires no global solve. Second, the adjoint displacement field $\boldsymbol{\lambda}^{(n)}$ is obtained from the linear problem \eqref{eq:4-13}.

The coefficient tensor of this linear problem is precisely the algorithmic tangent $\mathbb{D}^{ats\,(n)}$ (Equation~\eqref{eq:2-16}) at the converged Newton iteration of the forward analysis, and no new tangent needs to be assembled for the adjoint. In the implementation, $\mathbb{D}^{ats\,(n)}$ is reconstructed from $\theta^{(n)}$, $\bar\theta^{(n)}$, and $\hat{\boldsymbol{n}}^{(n)}$ obtained in the forward analysis and supplied as an anisotropic elasticity tensor of a linear elastic analysis. The right-hand side is the adjoint initial stress $\boldsymbol{\Sigma}^{il\,(n)}$ \eqref{eq:4-14}, which can be expressed within the framework of existing finite element formulations by prescribing it as an initial stress. The load originating from the objective function, $\mathbb{D}^e : \boldsymbol{f}_\sigma$, is injected only at the final step, as indicated by $\delta_{nN}$ in \eqref{eq:4-12}, and is transported to the subsequent steps through the chain of $\boldsymbol{P}$.

As a result, while the forward analysis carries the state pair $(\boldsymbol{\sigma}^{(n)}, \bar\alpha^{(n)})$ forward along the building direction, the adjoint analysis carries $(\boldsymbol{P}^{(n)}, \gamma^{(n)})$ in the reverse direction, in the identical ``initial stress + state variables'' structure. This is the implementation consequence of the mirror symmetry shown in Table~\ref{tab:4-1}.

Two points require attention in the implementation. First, the plastic gate $\chi_{\mathcal{P}}^{(n)}$ in the sensitivity expression \eqref{eq:4-17} and the branch of the $\boldsymbol{P}$ recurrence must be judged by the increment of the state variables stored by the forward analysis, $\Delta\bar\alpha^{(n)} = \bar\alpha^{(n)} - \bar\alpha^{(n-1)}$. A gate that reconstructs the trial stress a posteriori and repeats the yield check overlooks integration points that actually undergo plastic deformation in regions with large plastic deformation; the $\mathbb{D}^{ats}$ of such points is then treated as elastic, and the adjoint field $\boldsymbol{\lambda}$ is severely disturbed. Second, in the inactive domain $\Omega_I^{(n)}$, $\mathbb{D}^e = E_0 \cdot \mathrm{const.}$ does not depend on the design variable, and hence the explicit derivative with respect to $\bar z$ must be zero. The fact that term \circled{2} of the sensitivity expression \eqref{eq:4-17} takes the form of the ``pair of the stress and the adjoint strain'' automatically satisfies this requirement; in an expression rewritten to divide by the elastic modulus, $E_0$ cancels out and a spurious contribution of $\mathcal{O}(1)$ arises from the inactive steps.

\subsubsection{Optimization algorithm}\label{sec:5-1-3}

The optimization problem \eqref{eq:3-12} is solved by a gradient-based method. Once the sensitivities are available, standard sequential convex approximation methods are applicable; in this study, an MMA-type update rule that treats the volume constraint in an equality manner \cite{Svanberg1987,Svanberg2002} is used. The sharpness $\beta$ of the projection is increased geometrically with the iterations from its initial value to an upper bound. This continuation suppresses both the loss of gradient information in the early iterations and the intermediate densities in the final design. The overall optimization loop is presented in Algorithm~\ref{alg:3}.

\begin{algorithm}[tb]
\caption{Optimization loop for residual stress minimization}
\label{alg:3}
\begin{algorithmic}[1]
\State Give the initial design $\psi^{[0]}$; iteration counter $k = 0$
\Repeat
  \State Obtain the physical density $\bar z^{[k]}$ by the filter \eqref{eq:5-1} and the projection \eqref{eq:3-2} with parameter $\beta^{[k]}$
  \State Perform the layer-by-layer elastoplastic AM analysis (Algorithm~\ref{alg:1}) and store $\{\boldsymbol{\sigma}^{(n)}, \bar\alpha^{(n)}\}_{n=1}^{N}$
  \State Solve the static analysis \eqref{eq:3-6} of the final use to obtain $\hat{\boldsymbol{u}}$
  \State Evaluate the objective function $J$ \eqref{eq:3-10} and the constraints $l(\hat{\boldsymbol{u}})$ and $\bar V$
  \State Obtain $\{\boldsymbol{\lambda}^{(n)}, \boldsymbol{P}^{(n)}, \gamma^{(n)}\}$ by the reverse sweep of the adjoint (Algorithm~\ref{alg:2})
  \State Evaluate the design sensitivities \eqref{eq:4-17} and \eqref{eq:4-19} and transform them into $dJ/d\psi$ and $dg/d\psi$ by the chain \eqref{eq:5-2}
  \State Update $\psi^{[k+1]}$ by the sequential convex approximation
  \State Update $\beta^{[k+1]}$ by continuation up to the upper bound $\beta_{\max}$; $k \leftarrow k + 1$
\Until{the change of the design variables falls below the convergence tolerance or the maximum number of iterations is reached}
\State Output: optimized design $\bar z^{[k]}$
\end{algorithmic}
\end{algorithm}

Both the forward analyses on lines 4 and 5 and the adjoint analysis on line 7 are built on the weak-form interface of a general-purpose finite element software, and the filter, projection, sensitivity chain, and design variable update on lines 3, 8, and 9 are driven by external scripts. With the construction of Section~\ref{sec:5-1-2}, in which the algorithmic tangent $\mathbb{D}^{ats\,(n)}$ is supplied as an anisotropic elasticity tensor and the adjoint initial stress $\boldsymbol{\Sigma}^{il\,(n)}$ as an initial stress, the adjoint analysis is realized with the functionality of linear elastic analysis alone, and no dedicated adjoint solver needs to be implemented. In this study, COMSOL Multiphysics is used for the finite element analyses, and MATLAB is used to drive the optimization loop through LiveLink for MATLAB.

\subsection{Sensitivity verification by finite differences}\label{sec:5-2}

\subsubsection{Verification settings}\label{sec:5-2-1}

The validity of the sensitivity expression \eqref{eq:4-17} is verified by directly comparing the sensitivity with respect to the physical density, $dJ/d\bar z$, with central finite differences. Since the chain \eqref{eq:5-2} through the filter and projection is standard, it is excluded, and the core of the sensitivity, $dJ/d\bar z$, is verified alone.

\begin{figure}[pos=!htbp]
  \centering
  \includegraphics[width=\linewidth]{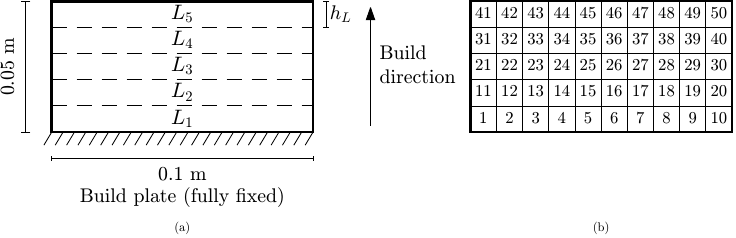}
  \caption{Analysis model used for the sensitivity verification. (a) Geometry, boundary conditions, and layer division ($N = 5$, $h_L = 0.01$ m). The inherent strain $\boldsymbol{\varepsilon}^*_0$ is applied only to the newly activated layer at each step, and no external load acts. (b) Element numbers ($10 \times 5 = 50$ elements).}
  \label{fig:5-1}
\end{figure}

The verification model is shown in Fig.~\ref{fig:5-1}. A two-dimensional plane stress rectangular domain of $0.1 \times 0.05$ m is divided into a $10 \times 5$ structured grid of 50 elements and into $N = 5$ layers of thickness $h_L = 0.01$ m along the building direction; that is, each layer corresponds to one row of elements. The elements are numbered from 1 at the left end of the bottom layer, increasing from 1 to 10 in the direction orthogonal to the building direction and then from 11 to 50 toward the upper layers (Fig.~\ref{fig:5-1}(b)). The bottom edge $\Gamma_u$ is fixed as the build plate, and no external load is applied; the stress originates solely from the inherent strain. The material properties are $E_1 = 72$ GPa, $\nu = 0.3$, $\sigma_{y1} = 243$ MPa, and $H_1 = 2.171$ GPa, and the floored power-law interpolation \eqref{eq:3-4} is used for the yield stress and the hardening modulus, with the same exponent as in the numerical examples of Section~\ref{sec:6}. The inherent strain is the design-independent uniaxial tensor $\boldsymbol{\varepsilon}^*_0 = (\varepsilon^*_1, 0, 0)$ with $\varepsilon^*_1 = -0.004$. The objective function is the unrelaxed P-norm with $p = 2$, and the baseline design is a uniform distribution of $\bar z = 0.9$. When the interpolation exponent of the yield stress is less than one, lower uniform densities remain elastic against the interpolated yield stress, and hence an operating point at which plastic deformation occurs is chosen. Furthermore, at $\bar z = 0.9$ the interpolation renders the yield strain larger than that of the solid material, and the number of layers is small, so that with $\varepsilon^*_1 = -0.003$ of Section~\ref{sec:2-5} the plastic deformation would localize in the four elements at the edges of the build plate. In this setting, the constraint imposed by the build plate causes all the ten elements of the bottom layer and two elements of the second layer (elements 11 and 20) to undergo plastic deformation; at the final step, 34 of the 450 Gauss points have deformed plastically, and the increment of the equivalent plastic strain is at most $5 \times 10^{-3}$. The region in which plastic strain remains at the completion of the build occupies 10.7\% of the domain, comparable to the 12.3\% of the numerical example of Section~\ref{sec:2-5}. In this mixed plastic/elastic state, all the path-dependent components of the formulation, namely the plastic term \circled{3}, the reverse chains of $\boldsymbol{P}$ and $\gamma$, and the transport by the $\mathbb{T}^{(n)}$ projections, are active.

The numerical sensitivity used for the comparison is evaluated by the central finite difference with respect to the physical density of each element:
\begin{equation}\label{eq:5-3}
\left. \frac{dJ}{d\bar z_e} \right|_{\mathrm{FD}}
= \frac{J\bigl(\bar{\boldsymbol{z}} + \Delta\bar z\, \boldsymbol{e}_e\bigr) - J\bigl(\bar{\boldsymbol{z}} - \Delta\bar z\, \boldsymbol{e}_e\bigr)}{2\,\Delta\bar z},
\qquad e = 1, 2, \ldots, N_{el}
\end{equation}
where $\bar{\boldsymbol{z}} = (\bar z_1, \ldots, \bar z_{N_{el}})$ is the baseline design of uniform $\bar z = 0.9$, $\boldsymbol{e}_e$ is the unit vector whose $e$-th component alone is 1, and $\Delta\bar z = 10^{-4}$ is the perturbation. In \eqref{eq:5-3}, the density of one element at a time is perturbed by $\pm\Delta\bar z$, and the layer-by-layer elastoplastic analysis of all $N$ steps is re-solved each time to evaluate $J$. The comparison over all 50 elements therefore requires $2 N_{el} = 100$ forward analyses, whereas the adjoint method provides the sensitivities of all the elements with a single reverse sweep. Since the reliability of finite differences in elastoplastic analyses is governed by the convergence level of the nonlinear solver, the solver tolerance is tightened to $10^{-9}$. With the default tolerance, the convergence residual remaining in the objective value contaminates the finite differences. As a prerequisite, we confirmed in advance that the objective values of the adjoint model and the finite difference model agree to ten digits.

\subsubsection{Results}\label{sec:5-2-2}

\begin{figure}[pos=!htbp]
  \centering
  \includegraphics[width=\linewidth]{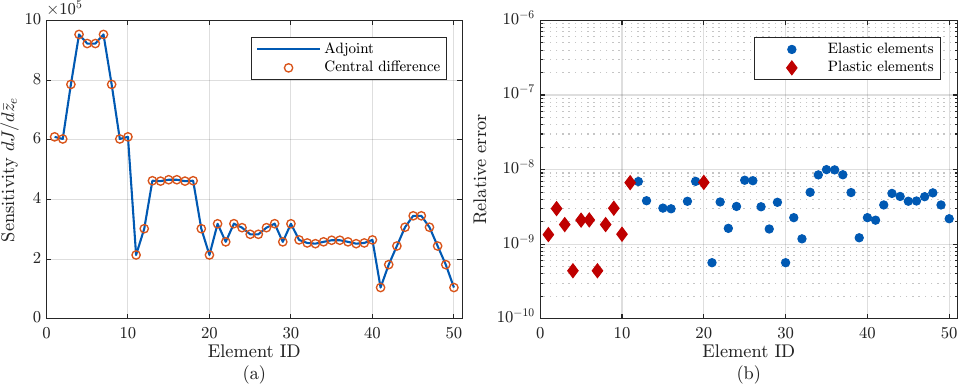}
  \caption{Element-wise comparison between the adjoint sensitivities and the central finite differences ($\Delta\bar z = 10^{-4}$, $\varepsilon^*_1 = -0.003$). (a) Overlay of the sensitivity distributions (solid line: adjoint; circles: central differences). (b) Relative error (circles: elastic elements; diamonds: plastically deformed elements).}
  \label{fig:5-2}
\end{figure}

The element-wise comparison is shown in Fig.~\ref{fig:5-2}. The relative error is defined as $|(dJ/d\bar z)_{FD} - (dJ/d\bar z)_{adj}| / \allowbreak |(dJ/d\bar z)_{adj}|$. As shown in Fig.~\ref{fig:5-2}(a), the adjoint sensitivities agree well with the central finite differences for all the elements. The relative error in Fig.~\ref{fig:5-2}(b) is below $10^{-8}$ for all the elements, and the agreement does not deteriorate in the plastically deformed elements marked by diamonds. These results demonstrate the correctness of the adjoint sensitivity formulation derived in Section~\ref{sec:4}.

\subsection{Computational complexity}\label{sec:5-3}

The computational cost per step of the adjoint analysis consists of one linear solve for the adjoint displacement field $\boldsymbol{\lambda}^{(n)}$ and the algebraic update of $(\boldsymbol{P}, \gamma)$ at each Gauss point. The former is a linear system with the same sparsity pattern as the Newton iterations of the forward analysis, at a cost of $\mathcal{O}(N_{dof}^{1.5})$, and the latter is $\mathcal{O}(N_{GP})$. The total over all the steps is therefore
\begin{equation}\label{eq:5-4}
\mathcal{O}\bigl(N \cdot N_{dof}^{1.5}\bigr)
\end{equation}
which is linear in the number of layers $N$. Since the forward analysis also requires $N$ steps, the cost of the sensitivity computation remains of the same order as the forward analysis.

This linearity is a consequence of constructing the adjoint while preserving the one-step recurrence structure of the state pair $(\boldsymbol{\sigma}^{(n)}, \bar\alpha^{(n)})$. In contrast, methods that expand the design dependence of the plastic strain explicitly by the chain rule \cite{DugastTo2023} must retain and update intermediate quantities of the form $\partial \boldsymbol{\varepsilon}^{p\,(n)} / \partial \boldsymbol{u}^{(j)}$ for each pair of steps $(n, j)$, resulting in $\mathcal{O}(N^2 \cdot N_{GP})$. The difference does not appear for problems with few layers, but this term becomes dominant for AM problems with a practical number of layers. The computation times reported in that reference for a 30-layer scale are interpreted as reflecting this scaling.

In addition to the complexity, we point out an implementation advantage. As described in Section~\ref{sec:5-1-2}, both the forward and adjoint analyses can be written in the identical structure of ``initial stress + carry-over of state variables.'' No additional state quantity such as the activation strain needs to be introduced for the adjoint, and no explicit recursion of the chain rule needs to be managed. This keeps the implementation cost low in that the existing finite element infrastructure can be used as it is, and it is also advantageous for parallelization in that the same partitioning strategy as the forward analysis can be applied.

In the next section, numerical examples in which the proposed method is applied to the residual stress minimization of a two-dimensional cantilever are presented.

\section{Numerical examples}\label{sec:6}

In this section, the residual stress minimization problem under a compliance constraint \eqref{eq:3-12} formulated in Section~\ref{sec:3} is applied to two design problems, a two-dimensional cantilever beam and a three-dimensional MBB beam, to validate the effectiveness of the proposed method. In both examples, as the basis of comparison, we use the optimized design of the minimum mean compliance problem \eqref{eq:3-8}, which takes no account of the residual stress during the building process and is hereafter called the reference design. The effect of the residual stress minimization is quantified by evaluating both designs with the identical elastoplastic layer-by-layer analysis of Section~\ref{sec:2}.

\subsection{Two-dimensional cantilever beam}\label{sec:6-1}

\subsubsection{Problem settings}\label{sec:6-1-1}

\begin{figure}[pos=!htbp]
  \centering
  \includegraphics[width=\linewidth]{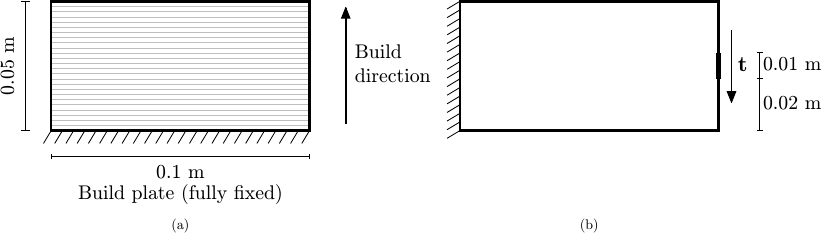}
  \caption{Problem settings of the two-dimensional example. (a) Building process: the bottom edge is fixed to the build plate, and $N = 25$ layers of thickness $h_L = 2$ mm are activated sequentially with the inherent strain $\boldsymbol{\varepsilon}^*_0$ applied (no external load). (b) Final use: the left edge is fixed, and a downward traction $\boldsymbol{t}$ is applied to the segment of height 10 mm at the center of the right edge.}
  \label{fig:6-1}
\end{figure}

The problem settings are shown in Fig.~\ref{fig:6-1}. The fixed design domain is a two-dimensional rectangular domain of width $100\,\mathrm{mm}$ and height $50\,\mathrm{mm}$ analyzed under plane stress. In the building process of Fig.~\ref{fig:6-1}(a), the bottom edge $\Gamma_u$ of the domain is fixed as the build plate, the domain is divided into $N = 25$ layers of thickness $h_L = 2\,\mathrm{mm}$ along the building direction, and the layer-by-layer elastoplastic analysis of Algorithm~\ref{alg:1} is performed. As in the verification of Section~\ref{sec:5}, the inherent strain is the design-independent uniaxial tensor $\boldsymbol{\varepsilon}^*_0 = (\varepsilon^*_1, 0, 0)$ with $\varepsilon^*_1 = -3.0 \times 10^{-3}$, applied only to the newly activated layer at each step. No external load acts. In the final use (Fig.~\ref{fig:6-1}(b)), the left edge is fixed as the support boundary $\Gamma_D$, and a downward traction $\boldsymbol{t} = (0, -1)\,\mathrm{MPa}$ is applied to the segment $\Gamma_t$ of height $10\,\mathrm{mm}$ at the center of the right edge, which constitutes a cantilever bending problem. Note that the build plate constraint $\Gamma_u$ in the building process differs from the support $\Gamma_D$ in the final use, that is, the build orientation does not coincide with the final-use orientation.

An aluminum alloy is assumed as the material, and the material properties are identical to those in Section~\ref{sec:2-5}. The analysis and optimization parameters are summarized in Table~\ref{tab:6-1}. The same value $0.5$ is used for the interpolation exponent $q_p$ of the yield stress and hardening modulus and for the relaxation exponent $q_s$ of the stress evaluation, as required in Sections~\ref{sec:3-1} and \ref{sec:3-3}, and the initial yield stress $\sigma_{y1}$ of the solid material is used as the normalization stress $\sigma_{\max}$.

\begin{table}[pos=!htbp]
  \centering
  \small
  \caption{Material, process, and optimization parameters of the two-dimensional example}
  \label{tab:6-1}
  \begin{tabularx}{\textwidth}{>{\raggedright\arraybackslash}p{2.5cm} >{\raggedright\arraybackslash}X >{\raggedright\arraybackslash}p{3.6cm}}
    \toprule
    Category & Symbol (definition) & Value \\
    \midrule
    Material & Young's modulus $E_1$ & $72$ GPa \\
             & Poisson's ratio $\nu$ & $0.3$ \\
             & Initial yield stress $\sigma_{y1}$ & $243$ MPa \\
             & Linear hardening modulus $H_1$ & $2.171$ GPa \\
    Process  & Inherent strain $\varepsilon^*_1$ & $-3.0 \times 10^{-3}$ \\
             & Layer thickness $h_L$ / number of layers $N$ & $2$ mm / $25$ \\
             & Young's modulus of quiet elements $E_0$ (Section~\ref{sec:2-2}) & $1$ kPa \\
    Interpolation / stress evaluation & RAMP \eqref{eq:3-3}: $q_E$, $\epsilon_E$ & $8$, $10^{-9}$ \\
             & Yield interpolation \eqref{eq:3-4}: $q_p$, $\epsilon_p$ & $0.5$, $10^{-3}$ \\
             & Stress relaxation \eqref{eq:3-9}: $q_s$ / normalization $\sigma_{\max}$ & $0.5$ / $\sigma_{y1}$ \\
             & P-norm \eqref{eq:3-10}: $p$ & $16$ \\
    Optimization & Filter radius \eqref{eq:3-1}: $r$ & $6$ mm \\
             & Projection \eqref{eq:3-2}: $\eta$ / $\beta$ & $0.5$ / $4 \to 16$ (geometric over 150 iterations) \\
             & Volume constraint: $V_{\max}$ & $0.5$ \\
             & Compliance constraint: $C_{\max}$ & $\lambda^{[k]} C_{\mathrm{ref}}$ (see below) \\
             & Initial design $\psi^{[0]}$ / max.\ iterations & uniform $0.5$ / $250$ \\
    \bottomrule
  \end{tabularx}
\end{table}

The domain is discretized with a $50 \times 25$ structured grid of $2 \times 2$ mm elements, and each layer corresponds to one row of elements. Elements with quadratic displacement interpolation are used for the layer-by-layer elastoplastic analysis and the adjoint analysis, and elements with linear displacement interpolation are used for the static analysis of the final use. The optimization is performed by Algorithm~\ref{alg:3}, and the convergence criterion is that the relative change of the objective function is below $10^{-3}$, the mean change of the design variables is below $3 \times 10^{-3}$, and all the constraints are satisfied, for three consecutive iterations after iteration 150.

The upper bound $C_{\max}$ of the compliance constraint is given on the basis of the reference design. First, the minimum mean compliance problem \eqref{eq:3-8} is solved under the identical domain, discretization, filter, and projection schedule to obtain the reference design of Fig.~\ref{fig:6-3}(a) and its mean compliance per unit thickness $C_{\mathrm{ref}} = 3.212 \times 10^{-2}\,\mathrm{N}$. In the residual stress minimization \eqref{eq:3-12}, the upper bound is set to $C_{\max} = \lambda^{[k]} C_{\mathrm{ref}}$ with a margin factor $\lambda^{[k]}$, which is decreased geometrically from $2.5$ to $1.2$ over 150 iterations with respect to the iteration $k$. This setting relaxes the constraint in the early iterations to secure the freedom of design change, and finally allows a compliance increase of up to 20\% relative to the reference design.

\subsubsection{Optimization results}\label{sec:6-1-2}

\begin{figure}[pos=!htbp]
  \centering
  \includegraphics[width=0.85\linewidth]{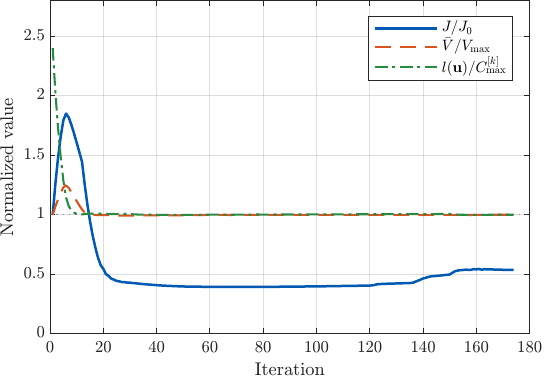}
  \caption{Convergence history of the optimization of the two-dimensional example: the objective function $J/J_0$ ($J_0$ is the value of the initial design), the volume constraint $\bar V/V_{\max}$, and the compliance constraint $l(\hat{\boldsymbol{u}})/C_{\max}^{[k]}$. Note that the upper bound $C_{\max}^{[k]} = \lambda^{[k]} C_{\mathrm{ref}}$ of the compliance constraint is tightened with the iterations ($\lambda = 1.2$ is reached at iteration 151).}
  \label{fig:6-2}
\end{figure}

The convergence history of the optimization is shown in Fig.~\ref{fig:6-2}. The initial uniform design largely violates the compliance constraint at $l/C_{\max} = 2.4$, and the optimization first drives both constraints to feasibility in about 15 iterations. The objective function temporarily increases during this process, then turns to decrease, and reaches its minimum $J/J_0 = 0.393$ at iteration 70, where $J_0$ denotes the objective value of the initial design. Thereafter, along with the continuation of the projection parameter $\beta$ and the progress of the compliance constraint ramp $\lambda^{[k]}$, which is completed at iteration 151, the objective function turns to increase, settles after the sharpening of the material boundary and the tightening of the constraints are completed, and the convergence criterion is satisfied at iteration 174. The objective of the final design is $J/J_0 = 0.536$, and both the volume and compliance constraints are active ($\bar V/V_{\max} = 0.9994$, $l/C_{\max} = 0.9995$).

\begin{figure}[pos=!htbp]
  \centering
  \begin{subfigure}[t]{0.49\linewidth}
    \centering
    \includegraphics[width=\linewidth]{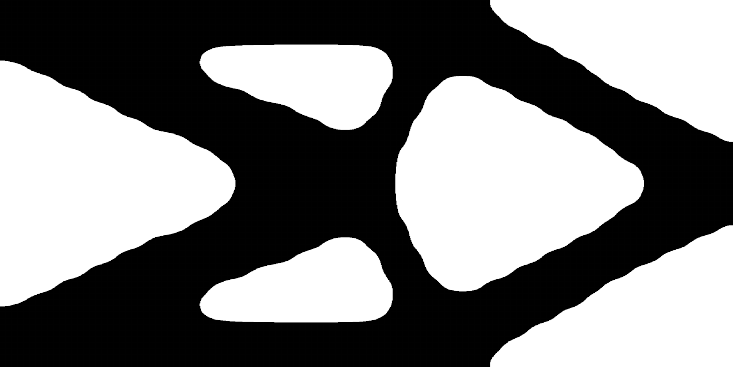}
    \caption{}
    \label{fig:6-3a}
  \end{subfigure}%
  \hfill%
  \begin{subfigure}[t]{0.49\linewidth}
    \centering
    \includegraphics[width=\linewidth]{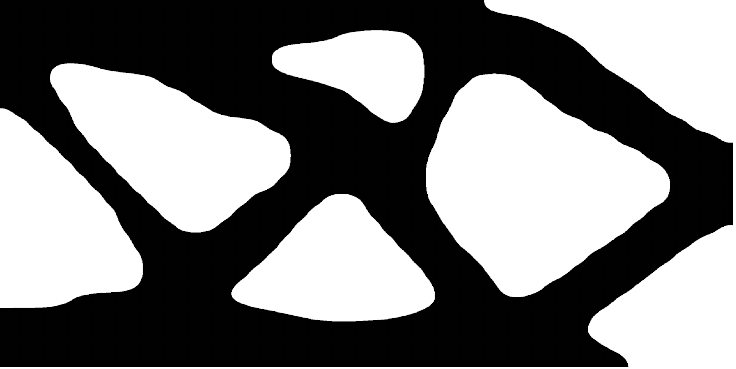}
    \caption{}
    \label{fig:6-3b}
  \end{subfigure}
  \caption{Optimized topologies of the two-dimensional example ($\phi \geq 0.5$; black = material). (a) Reference design (minimum mean compliance problem \eqref{eq:3-8}). (b) Proposed method (residual stress minimization \eqref{eq:3-12}).}
  \label{fig:6-3}
\end{figure}

The obtained optimized design is shown in Fig.~\ref{fig:6-3} together with the reference design. While the reference design in Fig.~\ref{fig:6-3}(a) is a cantilever truss composed of a few thick members, the design of the proposed method in Fig.~\ref{fig:6-3}(b) retains the bottom chord along the build plate and consists of a larger number of thinner diagonals and smaller openings.

\subsubsection{Evaluation and discussion of the residual stress}\label{sec:6-1-3}

\begin{figure}[pos=!htbp]
  \centering
  \begin{subfigure}[t]{0.49\linewidth}
    \centering
    \includegraphics[width=\linewidth]{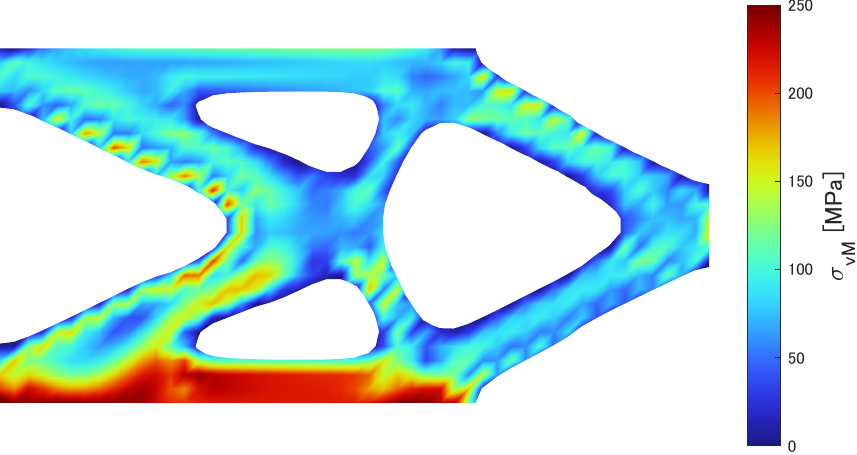}
    \caption{}
    \label{fig:6-4a}
  \end{subfigure}%
  \hfill%
  \begin{subfigure}[t]{0.49\linewidth}
    \centering
    \includegraphics[width=\linewidth]{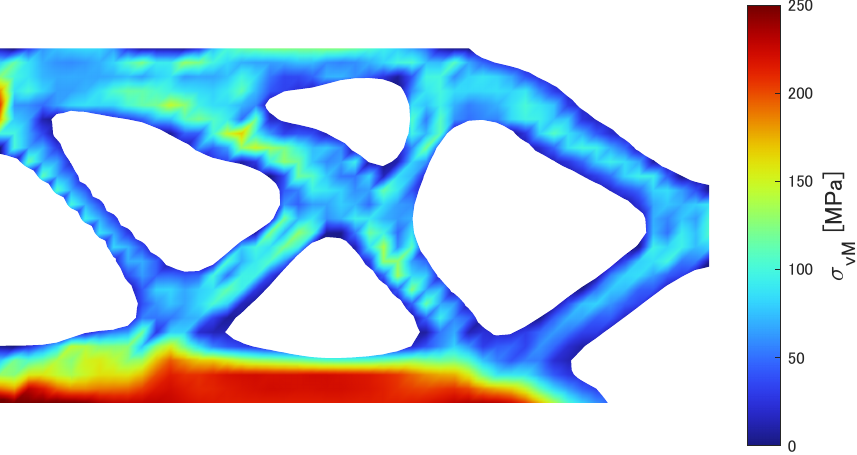}
    \caption{}
    \label{fig:6-4b}
  \end{subfigure}
  \caption{Distributions of the von Mises residual stress at the completion of the building process in the two-dimensional example. (a) Reference design. (b) Proposed method. The color range 0--250 MPa is common to both (values above the upper limit are saturated).}
  \label{fig:6-4}
\end{figure}

\begin{figure}[pos=!htbp]
  \centering
  \begin{subfigure}[t]{0.49\linewidth}
    \centering
    \includegraphics[width=\linewidth]{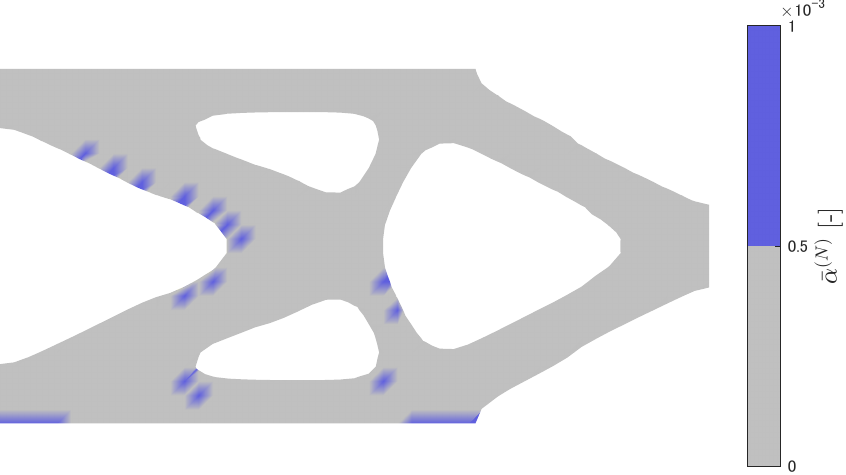}
    \caption{}
    \label{fig:6-5a}
  \end{subfigure}%
  \hfill%
  \begin{subfigure}[t]{0.49\linewidth}
    \centering
    \includegraphics[width=\linewidth]{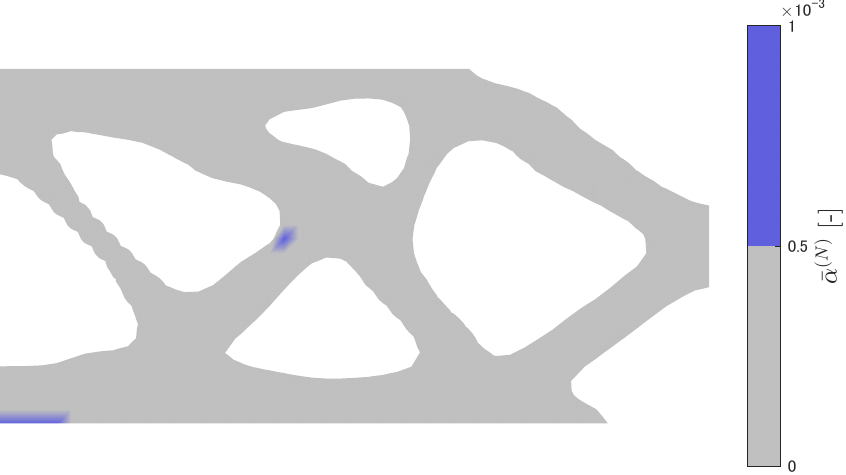}
    \caption{}
    \label{fig:6-5b}
  \end{subfigure}
  \caption{Distributions of the equivalent plastic strain $\bar\alpha^{(N)}$ at the completion of the building process in the two-dimensional example (blue: $\bar\alpha^{(N)} \geq 5 \times 10^{-4}$). (a) Reference design. (b) Proposed method.}
  \label{fig:6-5}
\end{figure}

Both designs are evaluated by the identical elastoplastic layer-by-layer analysis, and the residual stresses at the completion of the building process are compared. For the comparison between the designs, the von Mises stress without the relaxation \eqref{eq:3-9}, $\sigma_{vm}(\boldsymbol{\sigma}^{(N)})$, is evaluated at the Gauss points and aggregated over the solid region $\bar z \geq 0.5$. This is because the relaxed stress measure $\hat\sigma$ in \eqref{eq:3-9} is amplified at intermediate densities by $\eta_s/\eta_E > 1$ and is not suitable for comparing designs with different amounts of intermediate density. For the visualization in Figs.~\ref{fig:6-4} and \ref{fig:6-5}, the outlines are aligned with the topology plots of Fig.~\ref{fig:6-3} by using the $\phi = 0.5$ contour of a Helmholtz-type smoothing of the physical density field with a filter length of 2 mm. The evaluation results are summarized in Table~\ref{tab:6-2}.

\begin{table}[pos=!htbp]
  \centering
  \small
  \caption{Comparison of the residual stress at the completion of the building process in the two-dimensional example (von Mises stress without the relaxation \eqref{eq:3-9}, aggregated over the solid region $\bar z \geq 0.5$)}
  \label{tab:6-2}
  \begin{tabularx}{\textwidth}{>{\raggedright\arraybackslash}X >{\raggedright\arraybackslash}p{3.6cm} >{\raggedright\arraybackslash}p{3.6cm}}
    \toprule
    Measure & Reference design (min.\ compliance) & Proposed method (residual stress min.) \\
    \midrule
    Maximum $\sigma_{vm}^{\max}$ [MPa] & $268.6$ ($1.105\,\sigma_{y1}$) & $272.0$ ($1.119\,\sigma_{y1}$) \\
    P-norm ($p=16$, area-normalized) [MPa] & $199.0$ & $194.1$ \\
    Mean [MPa] & $103.5$ & $94.2$ \\
    Area fraction with $\sigma_{vm} \geq \sigma_{y1}$ & $0.0076$ & $0.0021$ \\
    Area fraction of the plastically deformed region & $0.023$ & $0.0067$ \\
    Maximum equivalent plastic strain $\bar\alpha^{(N)}$ & $2.03 \times 10^{-2}$ & $1.67 \times 10^{-2}$ \\
    Mean compliance $l(\hat{\boldsymbol{u}})$ [N] & $3.212 \times 10^{-2}$ & $3.853 \times 10^{-2}$ \\
    \bottomrule
  \end{tabularx}
\end{table}

The following observations are made from Table~\ref{tab:6-2}. First, the maximum residual stress hardly differs between the reference design and the proposed method, at 268.6 MPa versus 272.0 MPa. This is a consequence of the elastoplastic analysis model. The von Mises stress cannot exceed the yield surface $\sigma_y = \sigma_{y0} + H\bar\alpha$, and the constrained region near the build plate yields regardless of the design, so that the maximum stress is almost determined by the yield stress and the hardening. Indeed, the maxima of both designs exceed $\sigma_{y1}$ by 10--12\%, that is, they lie within the range of the yield surface expansion by hardening.

In contrast, the effect of the optimization clearly appears in the extent of the yielded region. The area fraction of the region subjected to stresses at or above the yield stress decreases from 0.0076 to 0.0021, and the area fraction of the plastically deformed region from 0.023 to 0.0067, both reduced by about 70\%. The maximum equivalent plastic strain also decreases by 18\%. The distributions in Fig.~\ref{fig:6-5} show the same tendency. In the reference design, numerous plastically deformed spots are distributed at the re-entrant corners of the bottom overhang and along the edges of the diagonals, in addition to the band-like zones at both ends of the build plate. In the proposed method, the plastic deformation shrinks to the band at the left end of the build plate and a single small spot at the crossing of the central diagonals; in particular, the bottom re-entrant corner at which the plastic deformation concentrated in the reference design is eliminated together with the structural feature itself. In other words, the optimization with the P-norm objective \eqref{eq:3-10} modifies the design so as to reduce the extent of the highly stressed and plastic strain accumulating regions, rather than the maximum stress value that is bounded by the yield surface. Since the accumulation of plastic strain during the building process is a direct cause of warpage and build failures, this reduction is a meaningful improvement from the viewpoint of manufacturing reliability. Note that the plastic deformation at the ends of the build plate originates from the boundary condition of the clamped foundation and cannot be removed by design changes.

This improvement comes at the price of stiffness. The mean compliance of the proposed method is $3.853 \times 10^{-2}\,\mathrm{N} = 1.20\, C_{\mathrm{ref}}$, which coincides with the constraint upper bound ($\lambda = 1.2$), and the constraint is active. That is, the residual stress reduction is achieved by using all of the allowed 20\% stiffness degradation, which shows that the trade-off between the stiffness and the residual stress reduction is explicitly controlled through the constraint in \eqref{eq:3-12}.

Finally, we comment on the numerical soundness of the analysis. All the plastically deformed Gauss points of the final design lie in the solid region ($\bar z > 0.9$), and no nonphysical plastic deformation occurs in the void, intermediate-density, or inactive regions. This is the effect of using the mild degradation with $q_p = q_s < 1$ and the residual value $\epsilon_p$ in the yield stress interpolation \eqref{eq:3-4} (Section~\ref{sec:3-1}), and it corroborates that the relaxed stress $\hat\sigma$ and the plastic response were evaluated consistently throughout the optimization.

\subsection{Three-dimensional MBB beam}\label{sec:6-2}

The second example applies the proposed method to a three-dimensional MBB beam to examine its applicability to three-dimensional problems, in which the inherent strain acts biaxially in the deposition plane and the plastic deformation during the building process is considerably more severe than in the two-dimensional example.

\subsubsection{Problem settings}\label{sec:6-2-1}

\begin{figure}[pos=!htbp]
  \centering
  \includegraphics[width=\linewidth]{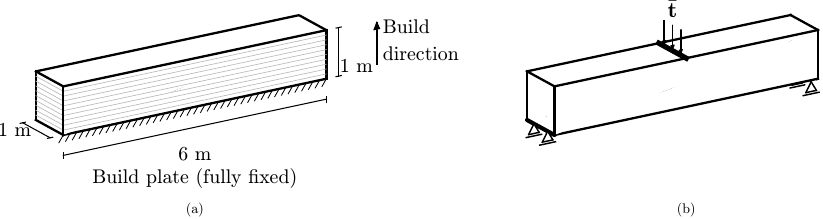}
  \caption{Problem settings of the three-dimensional MBB example (full domain; owing to the two vertical symmetries, only a quarter of the domain is analyzed, see the text). (a) Building process: the bottom face is fixed to the build plate, and $N = 10$ layers of thickness $h_L = 0.1$ m are activated sequentially with the in-plane biaxial inherent strain $\boldsymbol{\varepsilon}^*_0$ applied (no external load). (b) Final use: a downward line load $\bar{\boldsymbol{t}}$ is applied along the midspan line of the top face, and the bottom edges at both ends are supported by vertical rollers.}
  \label{fig:6-6}
\end{figure}

The problem settings are shown in Fig.~\ref{fig:6-6}. The fixed design domain of the MBB beam is a box of span $6\,\mathrm{m}$, depth $1\,\mathrm{m}$, and height $1\,\mathrm{m}$. In the building process (Fig.~\ref{fig:6-6}(a)), the bottom face of the domain is fixed as the build plate, the domain is divided into $N = 10$ layers of thickness $h_L = 0.1\,\mathrm{m}$ along the building direction, and the layer-by-layer elastoplastic analysis of Algorithm~\ref{alg:1} is performed. The inherent strain is the design-independent in-plane biaxial tensor $\boldsymbol{\varepsilon}^*_0 = (\varepsilon^*_1, \varepsilon^*_1, 0, 0, 0, 0)$ in Voigt notation with $\varepsilon^*_1 = -3.0 \times 10^{-3}$, which corresponds to the horizontal deposition plane of the three-dimensional process, applied only to the newly activated layer at each step; no external load acts. In the final use (Fig.~\ref{fig:6-6}(b)), a downward line load $\bar{\boldsymbol{t}} = 10\,\mathrm{kN/m}$ is applied along the midspan line of the top face across the depth, and the bottom edges at both ends of the span are supported by vertical rollers ($u_z = 0$), which constitutes the MBB bending problem. Taking advantage of the two vertical symmetries of the problem, symmetry conditions are imposed on the midspan plane and the mid-depth plane, and only a quarter of the domain, $3 \times 0.5 \times 1\,\mathrm{m}$, is discretized and analyzed in both the building-process and final-use analyses; accordingly, the integral quantities reported below, such as the volume and the mean compliance, refer to the quarter domain.

The material, the interpolation and stress-relaxation parameters, the P-norm exponent, the projection schedule, the compliance constraint ramp $\lambda^{[k]}$ ($2.5 \to 1.2$ over 150 iterations), and the convergence criterion are identical to those of the two-dimensional example (Table~\ref{tab:6-1}). Table~\ref{tab:6-3} summarizes the parameters changed in the three-dimensional example. The quarter domain is discretized with a $60 \times 10 \times 20$ structured grid of 12{,}000 cubic elements with edge length $0.05\,\mathrm{m}$, so that each layer corresponds to two element layers. The reference design is obtained by solving the minimum mean compliance problem \eqref{eq:3-8} under the identical discretization and schedules, and its mean compliance is $C_{\mathrm{ref}} = 3.746 \times 10^{-2}\,\mathrm{N \cdot m}$.

\begin{table}[pos=!htbp]
  \centering
  \small
  \caption{Parameters of the three-dimensional example (only those changed from Table~\ref{tab:6-1})}
  \label{tab:6-3}
  \begin{tabularx}{\textwidth}{>{\raggedright\arraybackslash}p{2.5cm} >{\raggedright\arraybackslash}X >{\raggedright\arraybackslash}p{4.8cm}}
    \toprule
    Category & Symbol (definition) & Value \\
    \midrule
    Process  & Inherent strain $\boldsymbol{\varepsilon}^*_0$ & in-plane biaxial, $\varepsilon^*_1 = -3.0 \times 10^{-3}$ \\
             & Layer thickness $h_L$ / number of layers $N$ & $0.1$ m / $10$ \\
    Discretization & Structured grid & $60 \times 10 \times 20$ (12{,}000 cubic elements, $h = 0.05$ m) \\
    Optimization & Filter radius \eqref{eq:3-1}: $r$ & $0.15$ m \\
             & Volume constraint: $V_{\max}$ & $0.4$ \\
             & Compliance constraint: $C_{\mathrm{ref}}$ & $3.746 \times 10^{-2}$ N$\cdot$m \\
    \bottomrule
  \end{tabularx}
\end{table}

\subsubsection{Optimization results}\label{sec:6-2-2}

\begin{figure}[pos=!htbp]
  \centering
  \includegraphics[width=0.85\linewidth]{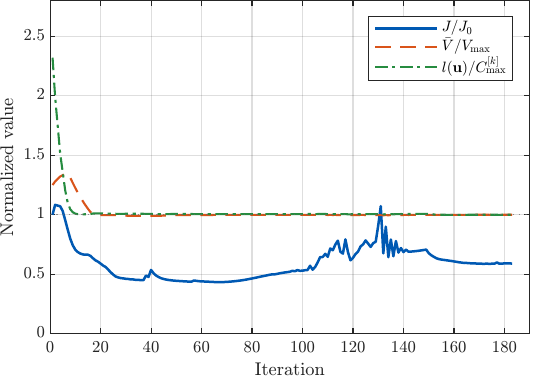}
  \caption{Convergence history of the optimization of the three-dimensional MBB example: the objective function $J/J_0$, the volume constraint $\bar V/V_{\max}$, and the compliance constraint $l(\hat{\boldsymbol{u}})/C_{\max}^{[k]}$ ($\lambda = 1.2$ is reached at iteration 151).}
  \label{fig:6-7}
\end{figure}

The convergence history is shown in Fig.~\ref{fig:6-7}. The initial uniform design violates both constraints ($\bar V/V_{\max} = 1.25$, $l/C_{\max} = 2.32$), and the optimization drives them to feasibility in about 15 iterations. The objective function reaches its minimum $J/J_0 = 0.433$ at iteration 67, turns to increase along with the continuation of $\beta$ and the tightening of $\lambda^{[k]}$, which is completed at iteration 151, and the convergence criterion is satisfied at iteration 183. The objective of the final design is $J/J_0 = 0.590$, and both the volume and compliance constraints are active ($\bar V/V_{\max} = 0.9996$, $l/C_{\max} = 0.9996$).

\begin{figure}[pos=!htbp]
  \centering
  \begin{subfigure}[t]{0.49\linewidth}
    \centering
    \includegraphics[width=\linewidth]{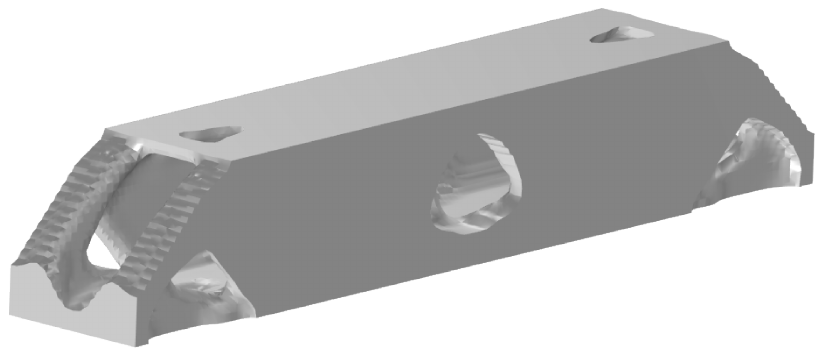}
    \caption{}
    \label{fig:6-8a}
  \end{subfigure}%
  \hfill%
  \begin{subfigure}[t]{0.49\linewidth}
    \centering
    \includegraphics[width=\linewidth]{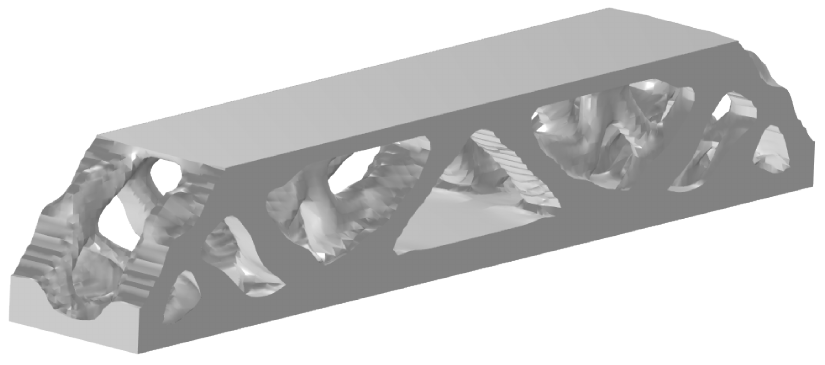}
    \caption{}
    \label{fig:6-8b}
  \end{subfigure}
  \caption{Optimized topologies of the three-dimensional MBB example (solid region $\bar z \geq 0.5$; the quarter-model solution is mirrored across the symmetry planes and the full geometry is shown). (a) Reference design (minimum mean compliance problem \eqref{eq:3-8}). (b) Proposed method (residual stress minimization \eqref{eq:3-12}).}
  \label{fig:6-8}
\end{figure}

The obtained optimized design is shown in Fig.~\ref{fig:6-8} together with the reference design. While the reference design in Fig.~\ref{fig:6-8}(a) concentrates the material into a deep arch-like structure with a few large openings, the design of the proposed method in Fig.~\ref{fig:6-8}(b) retains a continuous plate along the build plate and organizes the web above it into a larger number of thinner truss-like diagonals with smaller openings, which parallels the tendency observed in the two-dimensional example.

\subsubsection{Evaluation and discussion of the residual stress}\label{sec:6-2-3}

\begin{figure}[pos=!htbp]
  \centering
  \begin{subfigure}[t]{0.49\linewidth}
    \centering
    \includegraphics[width=\linewidth]{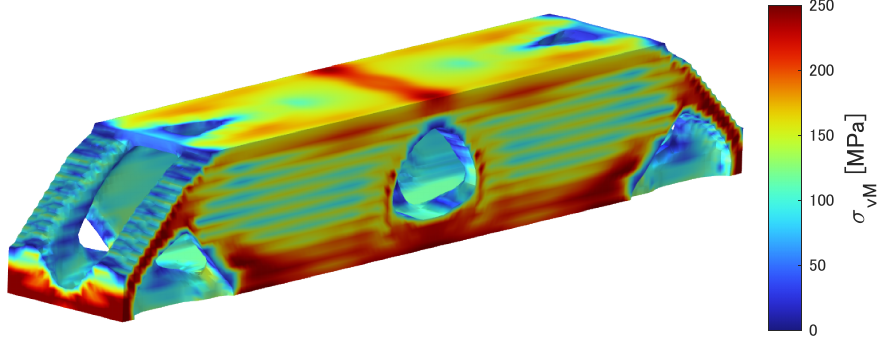}
    \caption{}
    \label{fig:6-9a}
  \end{subfigure}%
  \hfill%
  \begin{subfigure}[t]{0.49\linewidth}
    \centering
    \includegraphics[width=\linewidth]{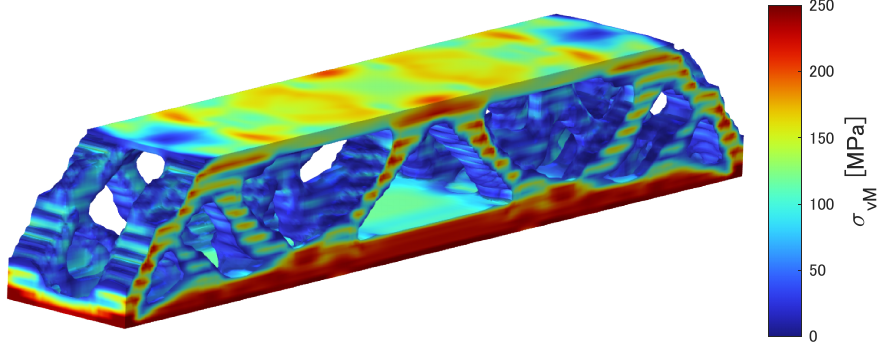}
    \caption{}
    \label{fig:6-9b}
  \end{subfigure}
  \caption{Distributions of the von Mises residual stress at the completion of the building process in the three-dimensional MBB example (full geometry shown as in Fig.~\ref{fig:6-8}). (a) Reference design. (b) Proposed method. The color range 0--250 MPa is common to both (values above the upper limit are saturated).}
  \label{fig:6-9}
\end{figure}

\begin{figure}[pos=!htbp]
  \centering
  \begin{subfigure}[t]{0.49\linewidth}
    \centering
    \includegraphics[width=\linewidth]{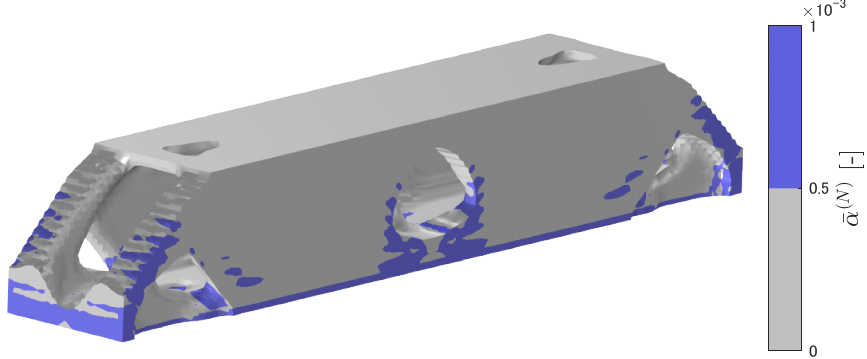}
    \caption{}
    \label{fig:6-10a}
  \end{subfigure}%
  \hfill%
  \begin{subfigure}[t]{0.49\linewidth}
    \centering
    \includegraphics[width=\linewidth]{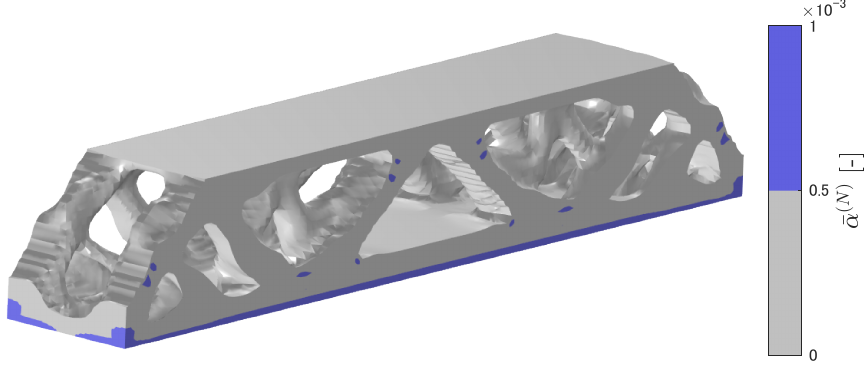}
    \caption{}
    \label{fig:6-10b}
  \end{subfigure}
  \caption{Distributions of the equivalent plastic strain $\bar\alpha^{(N)}$ at the completion of the building process in the three-dimensional MBB example (blue: $\bar\alpha^{(N)} \geq 5 \times 10^{-4}$; full geometry shown as in Fig.~\ref{fig:6-8}). (a) Reference design. (b) Proposed method.}
  \label{fig:6-10}
\end{figure}

Both designs are evaluated in the same manner as in Section~\ref{sec:6-1-3}, with the aggregation over the solid region using the volume weights of the quadrature. The results are summarized in Table~\ref{tab:6-4}, and the distributions of the residual stress and the equivalent plastic strain are shown in Figs.~\ref{fig:6-9} and \ref{fig:6-10}.

\begin{table}[pos=!htbp]
  \centering
  \small
  \caption{Comparison of the residual stress at the completion of the building process in the three-dimensional MBB example (von Mises stress without the relaxation \eqref{eq:3-9}, aggregated over the solid region $\bar z \geq 0.5$)}
  \label{tab:6-4}
  \begin{tabularx}{\textwidth}{>{\raggedright\arraybackslash}X >{\raggedright\arraybackslash}p{3.6cm} >{\raggedright\arraybackslash}p{3.6cm}}
    \toprule
    Measure & Reference design (min.\ compliance) & Proposed method (residual stress min.) \\
    \midrule
    Maximum $\sigma_{vm}^{\max}$ [MPa] & $291.7$ ($1.200\,\sigma_{y1}$) & $272.0$ ($1.119\,\sigma_{y1}$) \\
    P-norm ($p=16$, volume-normalized) [MPa] & $224.1$ & $219.9$ \\
    Mean [MPa] & $174.6$ & $138.9$ \\
    Volume fraction with $\sigma_{vm} \geq \sigma_{y1}$ & $0.127$ & $0.093$ \\
    Volume fraction of the plastically deformed region & $0.301$ & $0.203$ \\
    Maximum equivalent plastic strain $\bar\alpha^{(N)}$ & $2.24 \times 10^{-2}$ & $1.34 \times 10^{-2}$ \\
    Mean compliance $l(\hat{\boldsymbol{u}})$ [N$\cdot$m] & $3.746 \times 10^{-2}$ & $4.493 \times 10^{-2}$ \\
    \bottomrule
  \end{tabularx}
\end{table}

The following observations are made from Table~\ref{tab:6-4}. First, the plastic deformation in the three-dimensional example is considerably more severe than in the two-dimensional example: in the reference design, the volume fraction of the plastically deformed region reaches 0.301, more than ten times the corresponding area fraction 0.023 of the two-dimensional reference design. This is due to the in-plane biaxial inherent strain and the fully constrained bottom face, and it indicates that the three-dimensional example represents the severity of realistic building processes more faithfully.

Second, in contrast to the two-dimensional example, the maximum residual stress also decreases visibly, from 291.7 MPa to 272.0 MPa ($-6.8$\%). This reduction is still governed by the yield surface: for both designs, the maximum value agrees with $\sigma_{y1} + H_1 \bar\alpha^{(N)}_{\max}$ within rounding, that is, the maximum residual stress is attained at the most hardened point, and its reduction is a direct consequence of the reduction of the maximum equivalent plastic strain from $2.24 \times 10^{-2}$ to $1.34 \times 10^{-2}$ ($-40$\%). The extent of the highly stressed regions shrinks as in the two-dimensional example, and to a larger degree: the mean stress decreases by 20\%, the volume fraction of the region at or above the yield stress by 27\%, and the volume fraction of the plastically deformed region by 33\%. Fig.~\ref{fig:6-10} shows that the plastic deformation, which spreads over the arch and the bottom band in the reference design, is largely confined to the thin band just above the build plate in the proposed design. The improvement is again purchased by the allowed stiffness degradation: the mean compliance of the proposed method is $4.493 \times 10^{-2}\,\mathrm{N \cdot m} = 1.20\,C_{\mathrm{ref}}$, exactly at the constraint upper bound ($\lambda = 1.2$).

As in the two-dimensional example, no nonphysical plastic deformation occurs: all the plastically deformed Gauss points of both designs lie in the solid region, and the inactive (quiet) region remains strictly elastic throughout the building process.

In the next section, the conclusions of this paper and future work are presented.

\section{Conclusions}\label{sec:7}

In this paper, we constructed a topology optimization method for reducing the residual stress arising in metal additive manufacturing, which integrates a layer-by-layer process analysis model based on the elastoplastic inherent strain method with its exact adjoint sensitivity analysis.

First, in Section~\ref{sec:2}, a layer-by-layer elastoplastic analysis model without activation strains was presented. In this formulation, the incremental displacement is solved anew at each layer step, and the stress history is carried over by explicitly incorporating the stress accumulated up to the previous step into the constitutive equation. A newly activated layer thus carries no plastic or stress history and enters the analysis without any spurious initial stress, and the stress continuity across layer interfaces is guaranteed in the domain that has already been built. This formulation requires no correction term such as the activation strain that carries the displacement field at the moment of element activation over into the constitutive equation, and hence no additional dependencies between layer steps arise. In Section~\ref{sec:2-5}, a comparison with the linear elastic inherent strain method, in which only the plasticity is disabled under the identical layer division, quiet element treatment, and inherent strain loading, quantitatively showed that the linear elastic model overestimates the maximum residual stress by a factor of $4.6$ while the difference in the domain average remains 21\%, that is, the overestimation concentrates in the highly stressed regions.

Subsequently, in Section~\ref{sec:3}, the topology optimization problem \eqref{eq:3-12} that minimizes the P-norm of the residual stress at the completion of the building process under the volume and final-use compliance constraints was formulated based on the density method, and in Section~\ref{sec:4}, its design sensitivity was derived by the adjoint method. Taking the state as the pair of the stress and the equivalent plastic strain $(\boldsymbol{\sigma}, \bar\alpha)$ reduces the dependency between layer steps to a genuine one-step recurrence, and the adjoint fields can also be constructed as a layer-by-layer reverse sweep. Introducing the adjoint pseudo-stress $\boldsymbol{P}$, the adjoint equation of each reverse step reduces to a linear elasticity-type problem \eqref{eq:4-13} with the algorithmic tangent $\mathbb{D}^{ats}$ of the forward analysis as the stiffness and $\boldsymbol{\Sigma}^{il}$ as the initial stress, and it has the same structure as the forward equilibrium equation \eqref{eq:2-18}. The coefficient tensors obtained in the forward analysis can therefore be reused directly in the adjoint analysis, and neither additional state quantities for the adjoint nor explicit management of chain-rule recursions is needed. As a result, the total cost of the sensitivity computation remains at $\mathcal{O}(N \cdot N_{dof}^{1.5})$ \eqref{eq:5-4}, linear in the number of layers $N$, and of the same order as the forward analysis. This advantage over the existing method with $\mathcal{O}(N^2 \cdot N_{GP})$ \cite{DugastTo2023}, which expands the design dependence of the plastic strain explicitly by the chain rule, becomes more pronounced for problems with a practical number of layers. The derived sensitivities agreed with central finite differences at the $10^{-8}$ level for all the elements in Section~\ref{sec:5}, which confirmed the correctness of the formulation.

In Section~\ref{sec:6}, the proposed method was applied to a two-dimensional cantilever design problem and a three-dimensional MBB beam design problem. The two-dimensional optimization converged at iteration 174 with the objective function reduced to $0.536$ times that of the initial design, and the three-dimensional optimization, with 12{,}000 elements and 10 layers, converged at iteration 183 at $0.590$ times; in both cases the volume and compliance constraints were active. Evaluating the obtained designs and the reference designs of the minimum mean compliance problem, which takes no account of the residual stress, with the identical elastoplastic layer-by-layer analysis, the maximum residual stress hardly changed in the two-dimensional example, from $268.6\,\mathrm{MPa}$ to $272.0\,\mathrm{MPa}$. This is because the von Mises stress cannot exceed the yield surface and the constrained region near the build plate yields regardless of the design. On the other hand, the area fraction of the region subjected to stresses at or above the yield stress was reduced from $0.0076$ to $0.0021$, and that of the plastically deformed region from $0.023$ to $0.0067$, both by about 70\%, and the maximum equivalent plastic strain also decreased by 18\%. In the three-dimensional example, in which the in-plane biaxial inherent strain plastifies 30\% of the solid volume of the reference design, the same mechanism appeared more strongly: the maximum equivalent plastic strain decreased by 40\%, the volume fraction of the plastically deformed region by 33\%, and the maximum residual stress itself decreased from $291.7\,\mathrm{MPa}$ to $272.0\,\mathrm{MPa}$ in accordance with the reduced hardening. These improvements came at the price of a 20\% compliance increase, exactly at the constraint bound. That is, under the elastoplastic analysis model, the effect of the optimization with the stress P-norm objective appears as a reduction of the plastic strain accumulation and of the extent of the highly stressed regions, while the maximum value remains governed by the yield surface. This behavior cannot be captured by linear elastic models, in which the stress is not bounded by the yield surface, and it demonstrates the significance of incorporating the elastoplastic analysis model into the optimization.

Three directions are suggested for future work. The first is the integration with overhang constraints, which were not treated in this paper. Since the necessity of support structures directly affects the design freedom and the post-processing cost, it is desirable to incorporate geometric constraint methods \cite{Miki2023} into the present formulation and to extend it to a framework that simultaneously handles the physical manufacturing issue of residual stress and the geometric manufacturing issues. The second is the further scaling to high-resolution three-dimensional problems. The computational cost of the proposed method is linear in the number of layers; however, the computational cost of the three-dimensional elastoplastic layer-by-layer analysis itself remains large, and the combination with layer grouping, model reduction, and parallelization is a practical challenge for finer resolutions and larger numbers of layers. The third is the experimental validation of the analysis model and the optimized designs by comparison with measurements.

\printcredits

\section*{Declaration of competing interest}
The authors declare that they have no known competing financial interests or personal relationships that could have appeared to influence the work reported in this paper.

\section*{Data availability}
Data will be made available on request.

\section*{Acknowledgments}
This work was supported by the Japan Society for the Promotion of Science (JSPS) KAKENHI, Japan Grant Number JP24K07284.

\section*{Declaration of generative AI and AI-assisted technologies\\ in the manuscript preparation process}
During the preparation of this work the authors used Claude (Anthropic) in order to improve the language and readability of the manuscript. After using this tool, the authors reviewed and edited the content as needed and take full responsibility for the content of the published article.

\bibliographystyle{elsarticle-num}
\bibliography{refs}

\end{document}